\documentclass[letterpaper,twocolumn,10pt]{article}
\usepackage{usenix2019_v3}
\usepackage{verbatim}
\usepackage{tikz}
\usepackage{amsmath}
\usepackage{subfigure}
\usepackage{filecontents}
\usepackage{pifont}
\usepackage{algorithm}  
\usepackage{algorithmicx}  
\usepackage[noend]{algpseudocode}
\usepackage{booktabs}
\usepackage{upgreek}
\usepackage{array}
\newtheorem{theorem}{Theorem}

\newtheorem{corollary}{Corollary}
\newcommand*{\affaddr}[1]{#1} 
\newcommand*{\affmark}[1][*]{\textsuperscript{#1}}

\begin{document}

\date{}

\title{\Large \bf Programmable Cycle-Specified Queue  for Long-Distance\\ Industrial Deterministic Packet Scheduling}

\author{Yudong Huang\affmark[*], Shuo Wang\affmark[*], Shiyin Zhu\affmark[\dag], Guoyu Peng\affmark[*], Xinyuan Zhang\affmark[*], Tao Huang\affmark[*], Xinmin Liu\affmark[\dag]\\
	\affaddr{\affmark[*]State Key Laboratory of Networking and Switching Technology, BUPT, China}\\
	\affaddr{\affmark[\dag]H3C Technologies Co., Limited}\\
}
\maketitle

\begin{abstract}
The time-critical industrial applications pose intense demands for enabling long-distance deterministic networks. However, previous priority-based and weight-based scheduling methods focus on probabilistically reducing average delay, which ignores strictly guaranteeing task-oriented on-time packet delivery with bounded worst-case delay and jitter.  

 This paper proposes a new Programmable Cycle-Specified Queue (PCSQ) for  long-distance industrial deterministic packet scheduling. By implementing the first high-precision rotation dequeuing, PCSQ enables microsecond-level time slot  resource reservation (noted as $T$) and especially  jitter control of up to $2T$. Then, we propose the cycle tags computation to approximate cyclic scheduling algorithms, which allows packets to actively pick and lock their favorite queue in a sequence of nodes. Accordingly, PCSQ can precisely defer packets to any desired time. Further,  the queue coordination and cycle mapping mechanisms are delicately designed to solve the cycle-queue mismatch problem. Evaluation results show that PCSQ can schedule tens of thousands of time-sensitive flows and strictly guarantee $ms$-level delay and $\upmu$s-level jitter.
 
 
\end{abstract}

\section{Introduction}
The Industrial Internet of Things (IIoT) \cite{iiot} has brought a new era of network innovations\cite{URLLC}\cite{consumer_ser}. According to the ITU report\cite{ITUR}, machine-type devices  could reach 97 billion in 2030,  accompanied by emerging time-critical applications, such as smart grid\cite{smart_grid}, telediagnosis\cite{remote_sur}, remote control\cite{remote_control}\cite{tcpsbed}, digital twin\cite{pdt_rtrc}, and Metaverse\cite{metaverse}. However, unlike traditional best-effort transmission of bit information, industrial services require task-oriented on-time packet delivery with bounded delay and jitter. For instance, control loops depend on the timeliness of packet arrival\cite{network2030}; any misbehaved packets (e.g., early, late, out of order, or dropped) may cause severe production accidents and tremendous financial loss.


In response to the above challenges and trends, deterministic networking (DetNet)\cite{detnet_usecase}\cite{itut} has become a global research hotspot. The IEEE TSN\cite{intro_TSN} task group has put forward a family of standards, such as time synchronization and cyclic queuing and forwarding (CQF)\cite{802.1qch}, to support real-time transmission within the range of a factory or a local area, while the possibility of providing long-distance deterministic forwarding services for large-scale cyber-physical systems (e.g., networked control and haptic systems, interconnected multiple TSN fields) remains to be explored.


Ideally, on-time packet delivery means that packets can be precisely deferred\cite{damper}\cite{matters} for a relative or absolute amount of time at access nodes\cite{in_edge_control} or hop-by-hop\cite{csqf1}. In long-distance industrial scenarios, there are several challenges: (1) Dynamic flow contention. Since multiple flows will compete for the queue resources of output ports, the residence time is difficult to calculate and maintain\cite{mcqf}\cite{csqf_join}. Even though flow attributes are known prior, newly arriving flows may disrupt the performance of delivered flows. (2) Imperfect time synchronization\cite{no_ideal_clocks}\cite{tsn_peeper}. Due to the time drift of network devices, packets may miss the correct dequeuing duration,  causing the time slot resource reservation mechanism to fail. (3) Unnegligible link delay\cite{CENI}\cite{csqf_huang}. Long-distance link delays are hard to normalize and align with intra-node queuing behaviors, where inter-node collaboration for relative packet deferring should be carefully considered to alleviate traffic aggregations at downstream nodes.

On the one hand, cyclic scheduling is a promising paradigm to address the  above challenges. Specifically, cyclic scheduling divides the sending time of an output port into a series of equal time intervals; each time interval is called a cycle\cite{itp}. Packets are transmitted at a precise reserved cycle tagged in the packet headers. Hence, the end-to-end delay is bounded and predictable with the specific cycle information on each node along the path.  Many cyclic scheduling algorithms, such as Damper\cite{damper}, cycle specified queuing and forwarding (CSQF)\cite{csqf_huang}, and per packet value (PPV)\cite{ppv},  have been proposed to tackle the flow contention problem by mapping various flow features (e.g., periodic, sporadic, constant bit rate, and  arrival curve with committed burst size) to the underlying queue resources. Further, these high-level algorithms need the support of special underlying hardware  to overcome the non-ideal clocks and decouple the per-hop link delay from the queuing delay. Unfortunately, the effective hardware implementation is still lacking.

On the other hand, with the advantages of high throughput, flexibility, and customization, a hardware-programmable scheduler is indispensable to approximate cyclic scheduling algorithms. However, the existing programmable scheduling works, such as push-in-first-out (PIFO)\cite{PIFO}, push-in-extract-out (PIEO)\cite{PIEO}, and programmable calendar queues (PCQ)\cite{PCQ}, only consider simulating the priority-based and weight-based low-delay algorithms, which ignores guaranteeing the strict on-time delivery for industrial real-time traffic. Essentially, the rank computation in these works is restricted by local ordering, while the end-to-end properties, such as bounded delay and jitter, require the centralized admission control with global ordering. The push-in-pick-out (PIPO)\cite{PIPO} can express existing TSN scheduling algorithms in local-area networks, but it is unsuitable for large-scale deterministic networks with long-distance links.

In this paper, we observe that: \textbf{(\romannumeral1)} The uncertain queuing delay, which is inevitably caused by flow bursts and aggregations, prevents time-sensitive flows and best-effort traffic from co-transmitting. \textbf{(\romannumeral2)} The cyclic scheduling can strictly bound the queuing delay by smoothing the bursts and discretizing the queue waiting time. Based on the investigation, we propose a new Programmable Cycle-Specified Queue (PCSQ) that can express the features of cyclic scheduling to enable long-distance industrial deterministic networks. By cascading the PCSQ scheduler and strict-priority (SP) scheduler, the time-division multiplexing tunnels with guaranteed service level agreements (SLAs) are successfully established, which isolate time-sensitive flows from best-effort statistical multiplexing (\S \ref{pcsqfr}). The main contributions of this paper are:
 
 $\bullet$  We implement the high-precision rotation dequeuing, where frequency synchronization and time compensation modules are designed to keep $ns$-level time accuracy. PCSQ enables $us$-level time slot resource reservation (noted as $T$) and especially  jitter control of up to $2T$ (\S \ref{hprd}).

 $\bullet$ We propose the cycle tags computation to approximate cyclic scheduling algorithms, which is like buying time-stamped tickets for time-sensitive flows. PCSQ allows the packet to actively pick and lock its favorite queue in a sequence of nodes, i.e., be deferred to any desired time (\S \ref{ctc}).
 

 $\bullet$ Further, the queue coordination is enhanced to solve the cycle-queue mismatch problem (such as timeslot miss and traffic incast) in the device (\S \ref{qcn}).  The cycle mapping mechanism is presented to learn the cyclic scheduling behavior between long-distance connected nodes (\S \ref{cmbn}). 

 $\bullet$ To demonstrate the feasibility of our hardware design of PCSQ scheduler, we prototype it on an FPGA. The PCSQ-enabled switches can guarantee  bounded delay and jitter transmission over long-distance links on a realistic testbed, and can scale to tens of thousands of time-sensitive flows (\S \ref{evaluation}).

\section{Observations}

\begin{figure}[]
	\centering
	\setlength{\abovecaptionskip}{-0.4cm} 
	\includegraphics[width=3.4in]{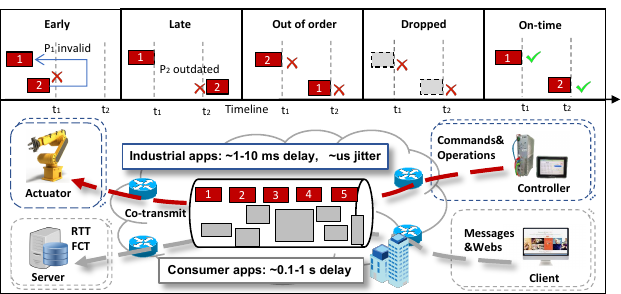}
	\caption{The on-time packet delivery is highly required to co-transmit the time-sensitive flows and best-effort traffic in typical industrial network scenarios.}
	\label{fig:on_time_delivery}
\end{figure}

\begin{table}[]
	\centering
	\small
	\setlength{\abovecaptionskip}{0.1cm} 
	\begin{tabular}{m{1.6cm}m{1.8cm}m{1.8cm}m{1.8cm}}
		\toprule[1pt]
		QoC & Delay (10 ms) & Jitter (2 ms) & Drops (15\%)  \\ \midrule[0.7pt]
		
		Kinematic-haptic loops & $\downarrow $ $\sim$88.2\% & $\downarrow $ $\sim$22.5\% & $\downarrow $ $\sim$49.25\% \\
		
		Kinematic-video loops & $\downarrow $ $\sim$6.1\% & $\downarrow $ $\sim$8.23\% & $\downarrow $ $\sim$59.3\% \\
		\bottomrule[1pt]
	\end{tabular}
	\caption{The network delay, jitter, and packet drops will significantly deteriorate the quality of control (QoC), compared to an ideal baseline for simulated network environments featuring 1 ms delay, zero jitter, and no packet drops.}
	\label{table3}
\end{table}

\begin{figure}[]
	\centering
	\setlength{\abovecaptionskip}{-0.4cm} 
	\includegraphics[width=3.4in]{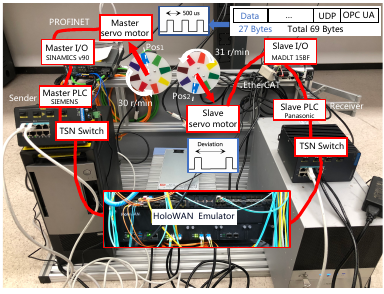}
	\caption{A testbed for TSN interconnection. The master PLC sends pulse signals across the emulated WAN to synchronize the position angle of the slave servo motor. $ \left| Pos_{1}-Pos_{2}  \right| $ is the measured metric of the position error.}
	\label{fig:plc_testbed}
\end{figure}

\begin{figure}[t]
	\centering
	\setlength{\abovecaptionskip}{-0.1cm} 
	\subfigcapskip=-10pt
	\subfigure[ Speed of 30 r/min.]{
		\begin{minipage}[t]{0.47\linewidth}
			\centering
			\includegraphics[width=\textwidth]{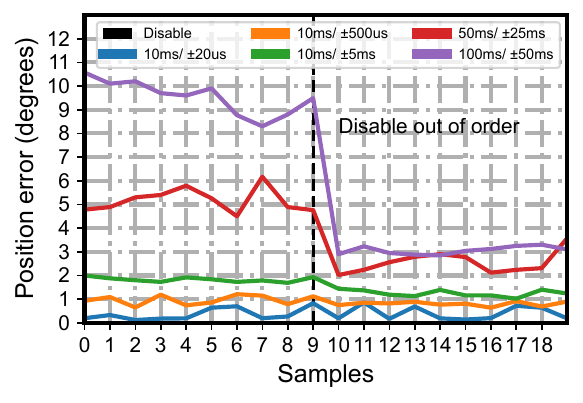}
			\label{fig20:jitter_30rpm}
		\end{minipage}
	}
	\subfigure[Speed of 150 r/min.]{
		\begin{minipage}[t]{0.47\linewidth}
			\centering
			\includegraphics[width=\textwidth]{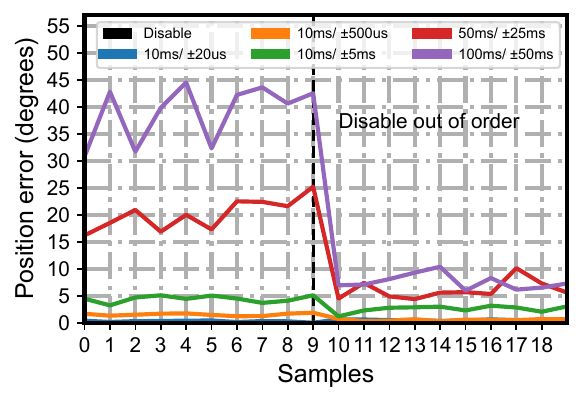}
			\label{fig20:jitter_150rpm}
		\end{minipage}
	}
	\centering
	\caption{ The impact of delay and jitter (delay/ $\pm$ jitter) on the remote control of servo motor. Out-of-order is disabled for the second half of samples. 
	}
	\label{fig:delay_jitter_speed}
\end{figure}

\subsection{Task-oriented On-time Packet Delivery }

\textbf{Features of industrial time-sensitive traffic:} Humans naturally can adapt to the congestion and collapse in the delivery of information, such as waiting for reconnections, retrying upon failures. Thus, network operators strive to reduce the average delays, e.g., round trip time (RTT) and flow completion time (FCT),  and tolerate the packet loss and considerable jitter. In contrast, machines execute operations strictly according to timelines. Task-oriented industrial applications require on-time arrival of command packets, e.g., packets $p_{1}$  must arrive before the time of $t_{1}$ and packet $p_{2}$ should arrive between  $t_{1}$ and  $t_{2}$ in Figure. \ref{fig:on_time_delivery}.  If packet $p_{2}$ arrives early (i.e., before  $t_{1}$), the command contained in packet $p_{1}$ will be invalid. Also, arriving late, out of order, and being dropped are not allowed. 

Specifically, time-critical applications send promised periodic/aperiodic traffic and require the bounded delay of milliseconds and  jitter of microseconds\cite{URLLC}\cite{in_edge_control}. For example, the Differential Protection traffic\cite{detnet_usecase} is just 64 kbps, but requires the maximum delay of less than 5 ms and the maximum jitter of less than 250 $\upmu$s. The remote control sends packets with sending period of 1-100 ms and packet size of 100-700 bytes, and the requirements are diverse as shown in Appendix \ref{appa}. For instance, the suturing operation in the da Vinci Surgical System\cite{gao2014jhu} issues kinematic packets every 10 ms\cite{tcpsbed}, updating Cartesian positions, orientations, velocities, angular velocities, and gripper angles to ensure real-time synchronized motion of the remote corresponding manipulators.

\textbf{The impact of network environments on QoC:} Literature \cite{qoc} shows that there is a strong correlation between quality of control (QoC) and end-to-end latency, jitter, and packet drops of network environments. We  briefly summarize part of the results in Table \ref{table3}. In the kinematic-haptic loops, the controller sends the kinematic packets and the actuator feedbacks the haptic information (e.g., vibration, pressure, smooth or rough, and soft or firm), where the QoC metrics could be the step disturbance, quadratic costs, and the rise time of the step response curves. In the kinematic-video loops, the feedback is the real-time video stream, and the QoC metrics could be positional errors and the maximum move speed. For instance, a visible positional error (>1 mm) can result cybersickness (e.g., dry eyes, headaches, dizziness, and nausea) that prevents users from long-term use of cyber-physical systems. It is also worth noting that terminal enhancement (such as delay prediction and trajectory prediction applications based on AI algorithms in cloud games\cite{zgaming}) cannot solve this problem, due to the real industrial physical properties.

\begin{figure}[t]
	\centering
	\setlength{\abovecaptionskip}{-0.1cm} 
	\subfigcapskip=-10pt
	\subfigure[ Speed of 30 r/min.]{
		\begin{minipage}[t]{0.47\linewidth}
			\centering
			\includegraphics[width=\textwidth]{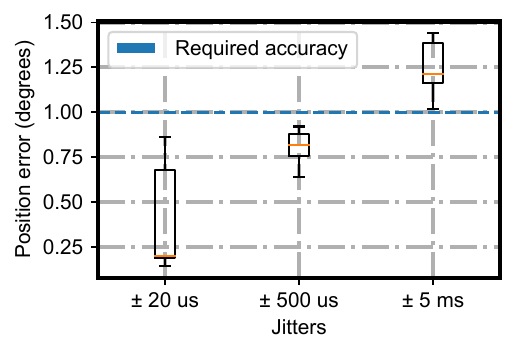}
			\label{fig20:jitter_box_30rpm}
		\end{minipage}
	}
	\subfigure[Speed of 150 r/min.]{
		\begin{minipage}[t]{0.45\linewidth}
			\centering
			\includegraphics[width=\textwidth]{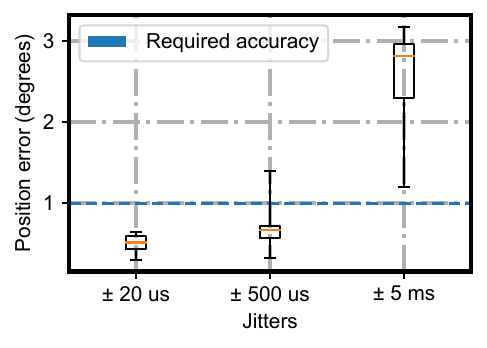}
			\label{fig20:jitter_box_150rpm}
		\end{minipage}
	}
	\centering
	\caption{ The impact of network jitter. The delay is set to 10 ms and out-of-order is disabled. 
	}
	\label{fig:jitter_box_speed}
\end{figure}

\begin{table}[t]
	\centering
	\small
	\setlength{\abovecaptionskip}{0.1cm} 
	\begin{tabular}{m{1.7cm}m{1.6cm}m{1.6cm}m{1.6cm}}
		\toprule[1pt]
		Position error (Max/Min/Avg) & Drops (10\%)& Drops (20\%)& Drops (30\%) \\ \midrule[0.7pt]
		
		30 r/min & 1.7/1.3/1.52& 16.3/2.3/7.46 & 23.8/5.4/11.46 \\
		
		150 r/min & 2.8/2.2/2.48 & 80.0/16.3/58.7  & 111.4/18.5/61.5 \\
		\bottomrule[1pt]
	\end{tabular}
	\caption{The impact of packet drops on the position error (degrees). 10\% drops is achieved by continuously discarding 100 of 1000 packets to simulate transient congestion.}
	\label{table20}
\end{table}

Furthermore, we built a TSN interconnection testbed to assess the impact of network parameters on master-slave motor synchronization. The servo motor is controlled by periodic pulse signals sent by the programmable logic controller (PLC). Each pulse drives the servo motor to rotate to a specific position with a default speed. Once a signal is disturbed or lost, the servo motor will automatically decelerate till stop running. The reaction time of the motor (from standstill to 3000 r/min) is within a few milliseconds. In Figure \ref{fig:plc_testbed}, the master PLC sends a 69-byte data packet, which contains the rotation speed and position information of the master servo motor, to the slave PLC every 500 $\upmu$s. The minimum static position alignment accuracy can be controlled at 0.001 degrees, and the required dynamic accuracy should be below 1 degree. 

As shown in Figure \ref{fig:delay_jitter_speed}, the maximum position error under 500 $\upmu$s jitter is 1.82 degrees, which exceeds the required 1 degree. In a common public network environment (100 ms delay, $\pm$50 ms jitter), the maximum position error expands to 10.57 degrees and 44.5 degrees at the speed of 30 r/min and 150 r/min, respectively. Moreover, we found that disabling out-of-order reduces the maximum position error to less than 3 degrees and less than 10 degrees.  In Figure \ref{fig:jitter_box_speed}, a jitter of 500 $\upmu$s alone still has a certain probability of making the accuracy exceed 1 degree when the speed is 150 r/min. Table \ref{table20} shows that packet loss is also unacceptable in industrial networks. About 20\% packet loss will cause systems to be out of work.

\begin{figure}[]
	\centering
	\setlength{\abovecaptionskip}{-0.4cm} 
	\includegraphics[width=3.4in]{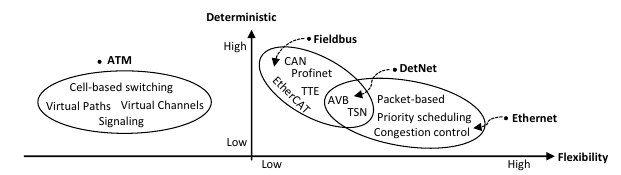}
	\caption{DetNet improves Ethernet's deterministic transmission capabilities and is more flexible than proprietary fieldbuses. Compared with Asynchronous Transfer Mode (ATM), DetNet reduces the technical complexity and has better compatibility with existing networks.}
	\label{fig:DetNet_compare}
\end{figure}

\begin{figure}[]
	\centering
	\subfigcapskip=-10pt
	\subfigure[ Long-tail delay.]{
		\begin{minipage}[t]{0.3\linewidth}
			\centering
			\includegraphics[width=\textwidth]{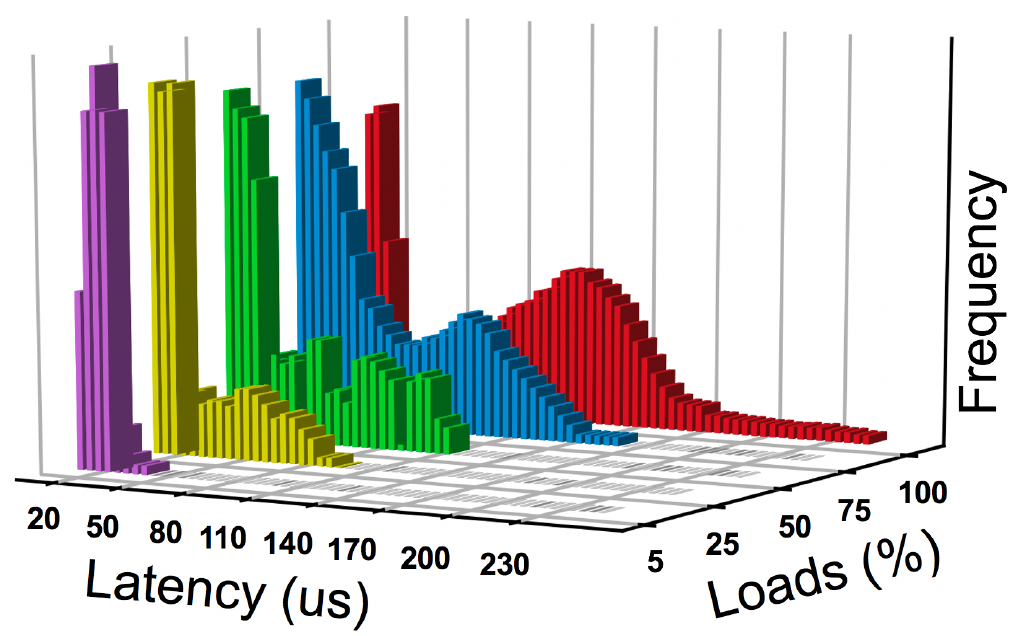}
			\label{fig:long_tail}
		\end{minipage}
	}
	\subfigure[Bursts.]{
		\begin{minipage}[t]{0.3\linewidth}
			\centering
			\includegraphics[width=\textwidth]{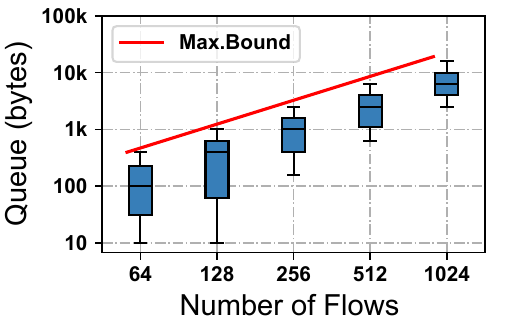}
			\label{fig:p1}
		\end{minipage}
	}
	\subfigure[ Aggregations.]{
		\begin{minipage}[t]{0.3\linewidth}
			\centering
			\includegraphics[width=\textwidth]{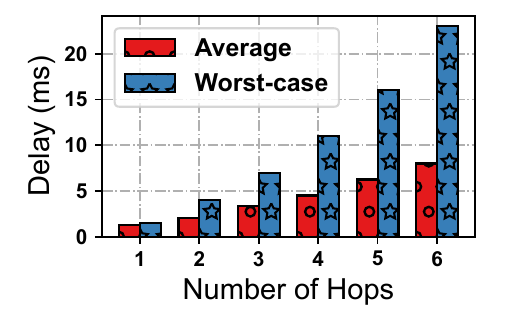}
			\label{fig:p2}
		\end{minipage}
	}%
	
	\centering
	\caption{ Evaluation results for best-effort forwarding. 
	}
\end{figure}

\textbf{Limitations of best-effort forwarding:} DetNet targets to bridge the gap between information technology (IT) and operational technology (OT) by enhancing real-time capabilities based on the Ethernet standard, as exhibited in Figure \ref{fig:DetNet_compare}. Traditional Ethernet/IP adopts the  best-effort forwarding that has the long-tail delay effect. To investigate this effect, we inject many token bucket flows into a commercial 10G Ethernet switch and measure the packet latency and queue length. As shown in Figure \ref{fig:long_tail}, with traffic load increases, more and more flows experience longer tail delays. The jitter is already in the order of tens of microseconds even for a load of 25\%. When the load is 100\%, the delay is unbounded and some packets are dropped. A very lightly-loaded (e.g., 5\%)  network may yield a low latency and small jitter, but huge bandwidth waste is unbearable. 

Essentially, the uncertain queuing delay is the culprit causing the long-tailed effect. As the number of injected flows grows, the queue length of the output port increases dramatically as shown in Figure \ref{fig:p1}. And the maximum queue length is proportional to the flow number (indicated by the red line). The reason is that bursty flows may send multiple packets at once and packets from multiple flows may arrive at the same time, thus they contend for the output port and the packet queue builds up significantly. For example, if five flows arrive simultaneously, the last flow has to wait in the queue until the other four flows are completely transmitted, resulting in a worst-case queuing delay of four flows' serialization time. 

This problem can become worse with flow aggregations at the downstream node.  In Figure \ref{fig:p2}, as the number of hops increases, the average delay of the flow grows gradually, while the worst-case delay increases sharply. In theory, it has been proved in \cite{aggre_bound} that the delay bound $D_{max}$ in a network with aggregate class-based strict priority scheduling is:
$$D_{max}=\left\{\begin{matrix}
\frac{(\Delta +\tau )h}{1-(h-1)\alpha} \ ,& if \ \alpha <\frac{1}{h-1} \\ 
\infty \ , & others  
\end{matrix}\right.\text{,} \eqno{(1)}$$ 
where $\tau$ is the maximum serialization delay and $\Delta$ is the maximum intra-node processing jitter. Specifically, $D_{max}$  is strongly correlated with the maximum link utilization $\alpha$ and the number of hops $h$. When the $\alpha$ approaches or exceeds $1/(h-1)$,  there is no upper bound on the delay. That means, for a network with ten hops, as long as the link utilization approaches or exceeds 11\%, the delay has no upper bound.


\textbf{\textit{Observation 1: The uncertain queuing delay, which is caused by flow bursts and aggregations, prevents time-sensitive flows and best-effort traffic from co-transmitting.  }}

\subsection{The Trends of Cyclic Scheduling} 

The best-effort forwarding, i.e., transmitting packets as soon as possible, is not applicable to industrial networks. Marking time-sensitive traffic as the highest priority can isolate the impact of best-effort traffic and reduce the average queuing delay, but the queue resource contention among time-sensitive traffic of the same priority is still uncontrollable. 

\begin{figure}[]
	\centering
	\subfigcapskip=-10pt
	\subfigure[ ]{
		\begin{minipage}[t]{0.46\linewidth}
			\centering
			\setlength{\abovecaptionskip}{-0.2cm} 
			\includegraphics[width=\textwidth]{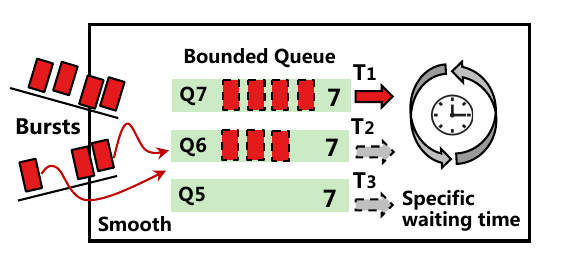}
			\label{fig:ob1}
		\end{minipage}
	}
	\subfigure[]{
		\begin{minipage}[t]{0.46\linewidth}
			\centering
			\setlength{\abovecaptionskip}{-0.2cm} 
			\includegraphics[width=\textwidth]{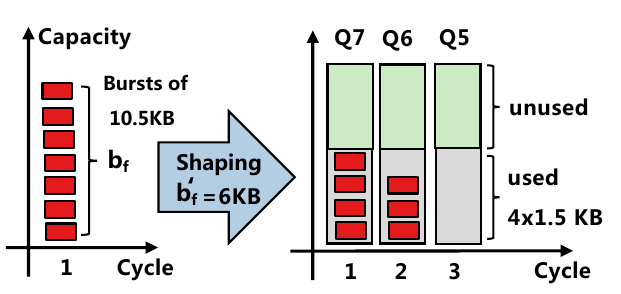}
			\label{fig:ob2}
		\end{minipage}
	}
	\centering
	\caption{ The cyclic scheduling can smooth the bursts $b_{f}$ of 10.5 KB with the shaping parameter $b^{'}_{f}$ of 6 KB.
	}
\end{figure}

As depicted in Figure \ref{fig:ob1}, we observe that the key point to bound the queuing delay is to restrict the maximum queue length at each output port and to bind the queue length to a time factor of the cycle $T$. During a cycle, packets in a queue must all be transmitted out from the current node.  Compared to previous active queue management (AQM) mechanisms and congestion control mechanisms that also have a threshold for single queue length, cyclic scheduling refines time slot allocation by changing the queue model from time-continuous to time-discrete\cite{sharp_edge}, and smoothes bursts in a time division multiplexing manner.  For instance, assuming the shaping parameter $b^{'}_{f}$ is 6 KB, the incoming bursts $b_{f}$ of 10.5 KB is smoothed into two queues with the discretized shaping delay of no more than $\left \lceil  \frac{b_{f}}{b^{'}_{f}}\right \rceil \times T$, which is $2T$ in Figure \ref{fig:ob2}. The theoretical analysis for cyclic scheduling is presented in Appendix \ref{appb}.

Based on the idea of cyclic scheduling, plenty of mechanisms and algorithms have been proposed to facilitate the bounded delay and jitter transmission, such as stop-and-go\cite{SAG}, Damper\cite{damper}, CQF\cite{802.1qch}, CSQF\cite{csqf_huang}, LDN\cite{towards}, and PPV(Per Packet Value)\cite{ppv}. The recently proposed CSQF as a representative instance is presented in Appendix \ref{appc}. In response to above trends and benefits, it is necessary to implement a cyclic scheduling paradigm into devices for long-distance deterministic transmission.

\textbf{\textit{Observation 2: The cyclic scheduling can strictly bound the queuing delay by smoothing the bursts and determine the queuing delay by discretizing the queue waiting time.}}

\section{Motivation}

\subsection{Packet Scheduling Dilemmas}\label{psd}
Several works concentrate on the practical experiments of small-scale deterministic networks (e.g., inside room \cite{tsn_ex}, a lab \cite{pi}, or a building \cite{tsn_opcua}), but no extensional studies are carried out in large-scale deterministic network devices where network services suffer from long-propagation delay and imperfect time synchronization.  Remarkably, programmable packet scheduling simplifies the testing and deployment of new scheduling algorithms, which is the best candidate for achieving cyclic scheduling.

Researchers have strived to find general scheduling primitives to cover as many scheduling algorithms as possible in the past decades. However, most of them target priority-based low-latency and weight-based fair scheduling algorithms, while the possibility of bounded delay and jitter scheduling has been largely ignored. Besides, it has been proved that no universal algorithm can
express all scheduling algorithms\cite{UPS}.  To deal with this dilemma, a novel programmable cycle-specified queue (PCSQ) scheduler is highly desirable to imitate cyclic scheduling. Next, we present the generic packet scheduling model, and detail the progressive relationship of PCSQ to other packet scheduling primitives.

\begin{figure}[]
	\centering
	\setlength{\abovecaptionskip}{-0.1cm} 
	\includegraphics[width=3.2in]{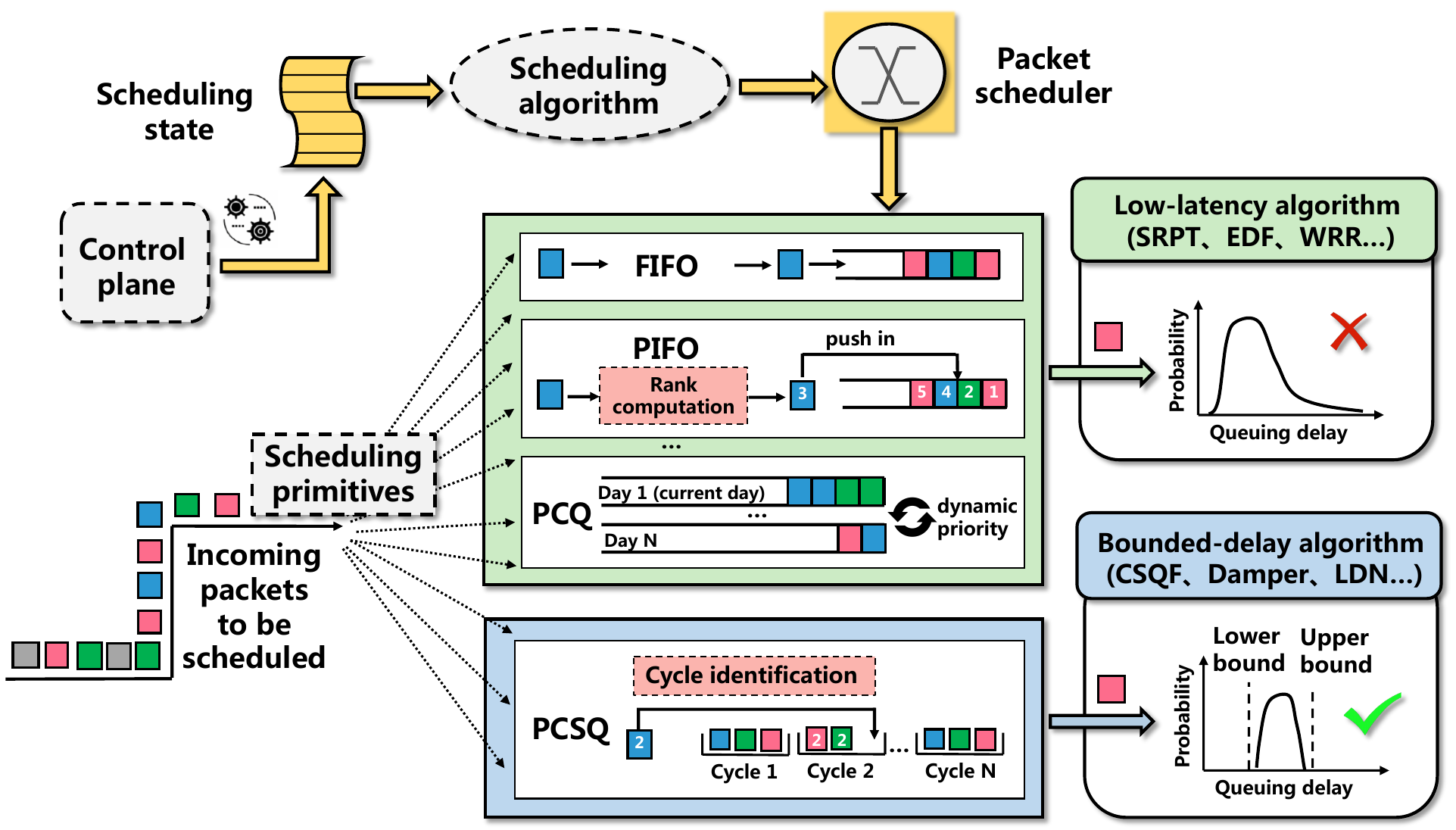}
	\caption{Packet scheduling primitives.}
	\label{fig:sch_pri}
\end{figure}

\textbf{Generic model:} In most Ethernet switches, the packet data is stored in a buffer pool when more than one input port is trying to send packets to the same output port simultaneously, and queues are used to store packets’ metadata for scheduling. As shown in Figure \ref{fig:sch_pri}, packets are selected  to enqueue, process, and dequeue according to some custom packet scheduling algorithm. The control plane can configure the scheduling state maintained in memory. And a packet scheduler is used to express and enforce the chosen scheduling algorithm.

\textbf{Scheduling algorithms:} Packet scheduling algorithm specifies when and in what order packets from each queue should be transmitted on the wire. The work-conserving algorithms do not let a link idle as long as there exists a packet waiting to be scheduled, such as Strict Priority (SP), Deficit Round Robin (DRR) \cite{DDR}, and Weighted Fair Queuing (WFQ)\cite{WFQ},  which are popular for their high bandwidth, low latency and fairness. The non-work conserving algorithms allow a link to be idle even when there are outstanding packets to send, such as Token Bucket (TB)\cite{TB} , TAS\cite{802.1qbv}, CQF\cite{tacq}, and CSQF\cite{load_balancing_csqf}\cite{csqf_join}. Since non-work conserving algorithms can specify the time to schedule a packet by  checking the eligible sending time against the current time,  they are naturally suitable for traffic shaping, such as rate-limiting and packet pacing, further carving the stringent lower bound and upper bound for delay and even jitter.

\textbf{Scheduling primitives:} We briefly introduce the five related primitives: FIFO, PIFO, PIEO, PIPO, and PCQ. The FIFO, i.e., first-in and first-out,  is the most basic scheduling primitive, which simply schedules packets in the order of their arrival. FIFO is easy to be implemented in hardware and multiple FIFOs can express the strict priority and some round robin algorithms, but it is incapable of expressing other popular scheduling algorithms. PIFO is a priority queue that associates a rank with each packet. The packet with the smallest rank (i.e., the highest priority) is always dequeued first. Different packet scheduling algorithms, such as Shortest Remaining Processing Time (SRPT)\cite{SRPT}, can be implemented on top of PIFO by changing the rank computation function.  PIEO not only utilizes the rank computation to enqueue packets,  but also allows dequeue from arbitrary positions by supporting predicate-based filtering at dequeue. PIPO is an approximation of PIEO that can express all existing TSN scheduling algorithms, while it is limited to local-area networks that require ideal clocks and negligible link delay. 

\textbf{The solution of PCSQ:} Programmable Calendar Queues (PCQ) is a recent proposal that supports the queue rotation mechanism.  It stores packets in a queue corresponding to a particular day and moves onto the next day after all packets are processed from the current day. This rotation mechanism allows scheduling algorithms to escalate the priorities of buffered packets with time and reuse emptied queues for incoming packets with low priority, which is close to the appearance of cyclic scheduling. Inspired by the PCQ, we envision that a  programmable cycle-specified queue (PCSQ) scheduler is ready to come out by jointly designing the priority queues and rotation queues. As shown in Figure \ref{fig:sch_pri}, in the PCSQ, packets are enqueued to a specific cycle-related queue by cycle identification, and dequeued in a time-discrete rotation manner with bounded queuing delay. To some extent, the cycle tags computation of PCSQ is like buying time-stamped tickets for time-sensitive flows, which allows the packet to actively pick and lock its favorite seat (queue) in a sequence of stations (nodes). While with the previous rank computation, packets can only passively choose to be received or dropped.

\subsection{Challenges for implementing PCSQ  }
It is significant to realize cyclic scheduling  based on programmable packet scheduling. However, the previous packet scheduling primitives, such as  FIFO, PIFO, PCQ, and PIPO,  cannot express the features of cyclic scheduling. There are mainly four challenges in implementing PCSQ. \textbf{(1)} PCSQ executes the enqueue operation by parsing the cycle tags in the packet header, rather than the rank calculation. \textbf{(2)} It requires multi-queue rotation similar to the dynamic priorities of calendar queues, but one for transmitting and the remaining for receiving. \textbf{(3)} PCSQ dequeues in a time-division multiplexing manner, requiring nanosecond-level frequency synchronization and microsecond-level cycle-based forwarding, which is completely different from the previous priority-based, packet-size-based, or deadline-based forwarding. Although frequency synchronization has been well studied, how to convert clocks to an arbitrary length of cycle and ensure the time accuracy of dequeuing is a significant new challenge. \textbf{(4)} Lastly and most importantly, traditional queue scheduling is asynchronous, which has the cycle-queue mismatch problem that packets at queue boundaries may miss specific time slots and disrupt subsequent cycle-based forwarding behavior, while synchronous scheduling may introduce additional resource overheads.

Motivated by the above problems, we aim to design 
the Programmable Cycle Specified Queue (PCSQ) primitive for implementing cyclic scheduling with the following properties:

\textbf{Microsecond-level dequeuing:} By utilizing frequency synchronization and timing compensation, PCSQ must enable high-precision rotation dequeuing to reserve microsecond-level time slot resources for time-sensitive flows.

\textbf{Bounded delay and low jitter:}  With the queue coordination and cycle mapping mechanism, the PSCQ must satisfy the queue-cycle match constraints and transmit flows to
the destination within the required bounded delay and low jitter.

\textbf{High scalability:} By exploiting segment routing, our scheme should have high scalability and schedule thousands of time-sensitive flows in large-scale deterministic networks.

We show how PCSQ achieves the first two goals in the next
section. The last goal is identified in the evaluation part.

\section{Packet Scheduling using  Programmable Cycle-Specified Queue }

\subsection{PCSQ Framework} \label{pcsqfr}

\begin{figure}[]
	\centering
	\setlength{\abovecaptionskip}{-0.5cm} 
	\includegraphics[width=3.4in]{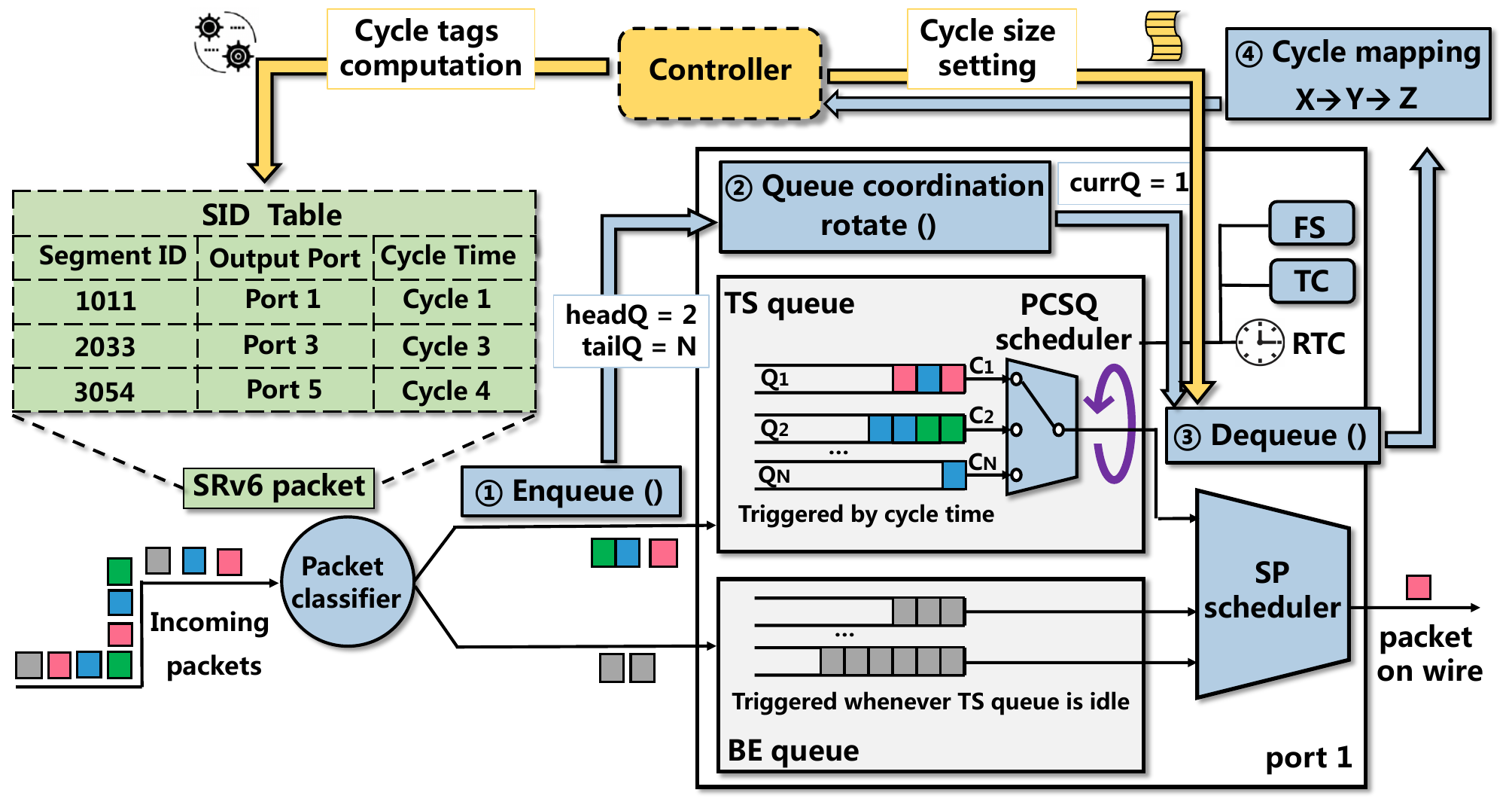}
	\caption{PCSQ Framework.}
	\label{fig:pcsq_framework}
\end{figure}

The PCSQ framework is depicted in Figure \ref{fig:pcsq_framework}. Switches have eight priority queues from seven to zero, and the highest priority is seven. We assume our PCSQ abstraction has a fixed number of buckets or FIFO queues, say N, each of which is marked with the highest priority to store the time-sensitive (TS) flows and mapped to an individual cycle time. The best-effort (BE) flows are marked with the remaining priority. The packet classifier distinguishes time-sensitive flows from best-effort flows based on QoS fields (e.g., DSCP or VLAN value) and sends them to the corresponding queue.

\textbf{The workflow of PCSQ:} Each time-sensitive packet is scheduled by making the following key decisions: \ding{172} First, the enqueue function identifies the port and cycle information by parsing the SRv6 header, and enqueues the packets to the specific queue binded to the cycle time. \ding{173} Second, the queue coordination function maintains the state of current transmitting queue (i.e., currQ=1) and receiving queue (i.e., headQ=2 and tailQ=N). The advancing of time can be done using a physical real-time clock (RTC), where the frequency synchronization (FS) module conducts the  frequency synchronization protocol and the timing compensation  (TC)  module converts the clocks to an arbitrary length of the cycle. Then,  the queue coordination function rotates the queue by periodically changing the  current transmitting queue to the next queue after a fixed time interval. \ding{174} Third, the packet is dequeued once the queue where the packet is located becomes the current transmitting queue. \ding{175} Finally, the cycle mapping function learns the cycle relationship between adjacent nodes, so the packet can be accurately sent to the queue of the next node with the next segment ID.

 To transmit time-sensitive flows and best-effort flows over one converged network, we cascade a strict-priority (SP) scheduler after the PCSQ scheduler. The BE queues are triggered whenever the TS queue is idle and the idle time is long enough to allow at least one current BE packet to be transmitted. Compared to the previous PIFO primitive, PCSQ scheduler needs to maintain states only at the granularity of switch queues (e.g., the queue corresponding to the current cycle). The scheduler does not require expensive sorting or comparisons to determine packet transmission order. 
 

\textbf{SLA configuration:} In general, Internet operators use leased tunnels\cite{b4} to ensure strict QoS. After users provide service level agreement (SLA) parameters such as committed bandwidth and priority to the operator, the service can be activated in seconds. PCSQ is in the same vein as leased tunnels and requires more parameters such as deadline and jitter. Ideally, the controller may directly interact with end devices and update the configuration in microseconds\cite{shoal}. Next, we separately present the key functions of dequeuing,  queue coordination, cycle mapping, and cycle tags computation.

\subsection{High-Precision Rotation Dequeuing} \label{hprd}
Traditional round-robin scheduling is based on bit counters and bandwidth calculations, such as WRR (Weighted Round Robin) and WFQ (Weighted Fair Queuing), which cannot provide strict guarantees on hop-by-hop queuing delays. The cyclic scheduling of PCSQ is based on the advancing of time, which arouses a new challenge of microsecond-level time slot calculation. For instance, if the cycle size is 10 $\upmu$s, the current transmitting queue needs to complete the dequeue operation within 10 $\upmu$s, and the  cycle size of 10 $\upmu$s at each node must be the same. PCSQ realizes high-precision rotation dequeuing with the  frequency synchronization (FS) and time compensation (TC) modules.

\textbf{Frequency synchronization:} Native Ethernet (IEEE 802.3) is asynchronous with no timing traceability to a reference clock, and there are two representative protocols for network-wide frequency synchronization: precision time protocol (PTP) and synchronous Ethernet (SyncE).  PTP\cite{ptp} is a data-link layer protocol that can achieve nanosecond-level accuracy under idle network conditions, but achieve only sub-microsecond level accuracy under network congestion. 

SyncE\cite{synce} is a physical layer frequency distribution mechanism, which requires an external line card to be attached to the Ethernet interface. The master node transmits the bit clock signal to slave nodes and the latter recovers the clock via the transceiver CDR (Clock Data Recovery).  SyncE has the advantage of relying on the physical layer only, thus the clock synchronization quality is not influenced by impairments introduced by the upper layers, such as packet delay variation (PDV), and packet losses\cite{synce_em}. While native Ethernet has an independent free-running  clock with frequency accuracy not exceeding $\pm$ 100 ppm (parts per million), SyncE  can achieve long-term accuracy of $\pm$ 10 parts-per-trillion as defined in ITU G.811\cite{synce}. Thus, this paper adopts the SyncE as the frequency synchronization mechanism.


\textbf{Timing compensation:} The controller can set the cycle size by modifying the localbus configuration register. And the time compensation module is responsible for converting clocks to an arbitrary cycle length. When the oscillator frequency of the device is a rational number (e.g., 400Mhz with the clock of 2.5 ns ) and the required cycle is also an integer multiple of the clock (e.g., 10 $\upmu$s), it is easy to obtain accurate time slot by iterative calculation (e.g., 2.5 ns for 4000 iterations is 10 $\upmu$s). However, when the device frequency is not a rational number, or the required cycle is not an integer multiple of the clock, a cycle conversion deviation will occur, which may further introduce a widening time-slot cumulative bias over the long run. To
 solve this problem,  we keep the cycle in nanoseconds to twelve decimal places and design a five-level cascade counter from $zs$, $as$, $fs$, $ps$ to $ns$. Our design can calibrate the conversion deviation and obtain ns-level accurate cycle time to achieve $\upmu$s-level rotation dequeuing.
\begin{figure}[]
	\centering
	\subfigcapskip=-10pt
	\subfigure[ Timeslot miss.]{
		\begin{minipage}[t]{0.46\linewidth}
			\centering
			\setlength{\abovecaptionskip}{-0.2cm} 
			\includegraphics[width=\textwidth, height=0.619\textwidth]{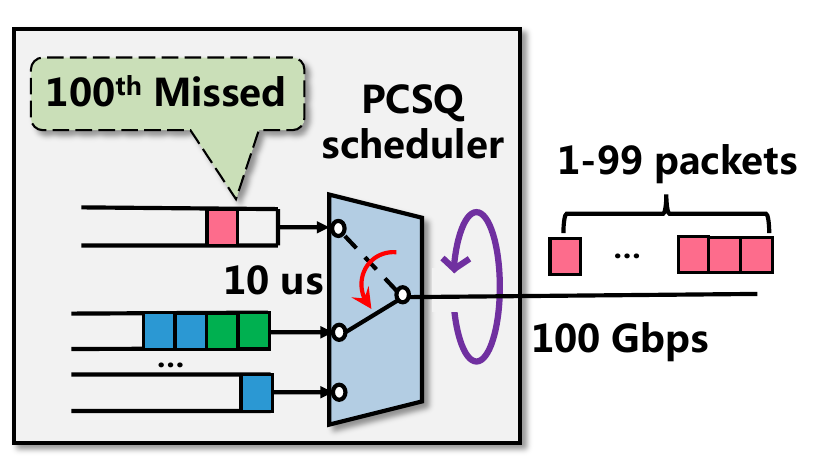}
			\label{fig:queue_coord1}
		\end{minipage}
	}
	\subfigure[Traffic incast.]{
		\begin{minipage}[t]{0.46\linewidth}
			\centering
			\setlength{\abovecaptionskip}{-0.2cm} 
			\includegraphics[width=\textwidth, height=0.619\textwidth]{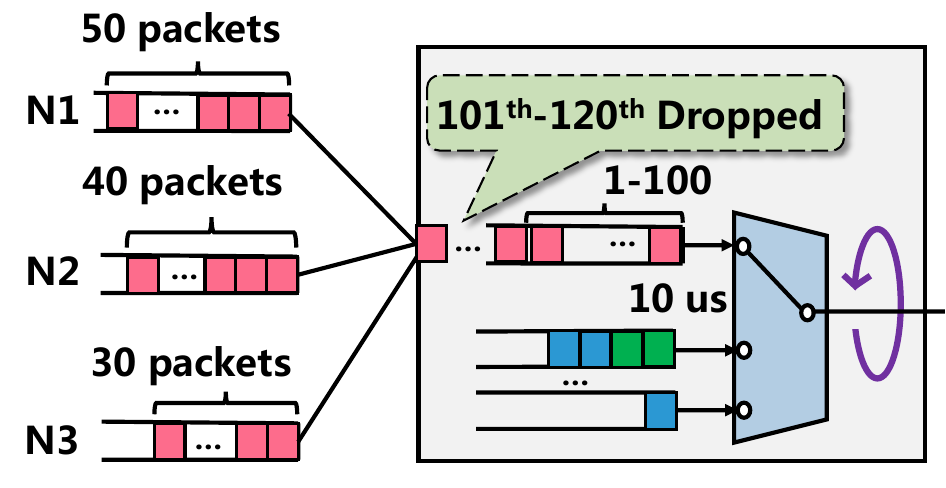}
			\label{fig:queue_coord2}
		\end{minipage}
	}
	\centering
	\caption{  Two cycle-queue mismatch cases.
	}
\end{figure}

\subsection{  Queue Coordination in Nodes} \label{qcn}

Since traditional queue scheduling is asynchronous, the enqueue and dequeue operations are separate and unaware. Asynchronous scheduling has the cycle-queue mismatch problem that packets at queue boundaries may miss specific time slots and disrupt subsequent cycle-based forwarding behavior. For the cyclic scheduling, when a rotation happens, we need to make sure all packets from the current transmitting queue are drained completely, which requires maintaining the state of head queue and notifying the enqueue operation when the dequeuing has been finished. 

We first analyze the two cycle-queue mismatch cases of timeslot miss and traffic incast. Then, to ensure that the cycle tag information in the packet header is consistent with the actual underlying queue scheduling behavior, we design the enhanced enqueue function and dequeue function for the queue coordination mechanism.

\textbf{Case-1: Cycle-queue mismatch caused by timeslot miss.} In this case, some packages at the tail of the queue may miss the current timeslot after the cycle time has elapsed and the head of the dequeuing has moved to the next queue. As shown in Figure \ref{fig:queue_coord1}, we assume that the port rate is 100 Gbps, the cycle size is set to 10 $\upmu$s,  and the maximum queue length is 100 packets. The $100^{th}$ packet may  miss the current timeslot and wait a total hyper-cycle until the located queue becomes the transmission queue again. One might argue that if the sum of the packet sizes does not exceed 125 KB (i.e., 100 Gbps$\times$10 $\upmu$s), then the packets can always be transmitted within 10 $\upmu$s. However, assuming that the last packet is 64 Bytes, the required transmission time is 0.00512 $\upmu$s, which only accounts for about five ten thousandths of 10 $\upmu$s. Therefore, if we only check whether the queue is full when packets enter the queue, but do not judge whether the queue is empty when the packets are dequeuing, it is possible that the queue has been rotated before the transmission is completed. Another way to deal with the timeslot miss problem is to over-reserve time slots, e.g., only enqueuing 80 packets even though one cycle time can transmit 100 packets. Nevertheless, over-reservation fails to exert the forwarding capability of the device, which violates the goal of maximizing the number of time-sensitive flows that can be scheduled.

\begin{algorithm}[t]
	\caption{PCSQ Enqueue}
	\begin{algorithmic}[1]

		\Function{ENQUEUE}{\textit{pkt}} 
		\State \textit{S} $\gets$ \textit{Queue.size};  \textit{N} $\gets$ \textit{Queue.num}
		\State \textit{sid} $\gets$ \textit{pkt.SID}
		\State \textit{qid} $\gets$ (\textit{currQ} + \textit{sid}) \% \textit{N}
		\If{\textit{qid} != \textit{currQ} $\&\&$ Queue(\textit{qid}).length < \textit{S}}  
		\State Queue(\textit{qid}).enqueue(\textit{pkt}) 
		\State Queue(\textit{qid}).length++
		\Else 
		\State \textit{qid} $\gets$ (\textit{currQ} + \textit{sid} + 1) \% \textit{N}
		\If{Queue(\textit{qid}).length < \textit{S}}  
		\State Queue(\textit{qid}).enqueue(\textit{pkt}) 
		\State Queue(\textit{qid}).length++
		\Else 
		\State Drop \textit{pkt}
		\EndIf
		\EndIf
		\EndFunction
	\end{algorithmic}
\end{algorithm}

\begin{algorithm}[t]
	\caption{PCSQ Dequeue and Rotation }
	\begin{algorithmic}[1]
		\Function{Dequeue and rotate}{} 
		\State  \textit{C} $\gets$  \textit{Cycle.size}; \textit{N} $\gets$ \textit{Queue.num}
		\State $t_{now} \gets$ \textit{RTC.currTime}; \textit{Flag} $\gets t_{now}$ 
		\While{true} 
			\While{$t_{now} $ $<$ \textit{Flag} +  \textit{C}}
			\If{Queue(\textit{currQ}) is not empty}
			\For {each \textit{pkt} in Queue(\textit{currQ})}
			\State \textit{pkt} $\gets$ Queue(\textit{currQ}).dequeue()
			\State Send \textit{pkt} 
			\EndFor
			\EndIf
			\State $t_{now} \gets$ \textit{RTC.currTime}
			\EndWhile
   \State \textit{Flag} $\gets t_{now}$ 
   \State \textit{currQ} $\gets$ (\textit{currQ} + 1) \% \textit{N} 
\EndWhile
		\EndFunction
	\end{algorithmic}
\end{algorithm}

\textbf{Case-2: Cycle-queue mismatch caused by traffic incast.} When a node has multiple upstream nodes, the queue forwarding behavior may not conform to the cycle tags due to traffic incast. As shown in Figure \ref{fig:queue_coord2}, the traffic of the three nodes arrives at the same time and starts to calculate the cycle tags. When viewing each of them, the queue length meets the demands. However, traffic aggregating to the same queue causes queue overflow, so all packets cannot be transmitted within 10 $\upmu$s and the 100$^{th}$ to 120$^{th}$ packets will be dropped. On the contrary, time-sensitive flows are critical traffic, which are not allowed to be dropped. To tackle the traffic incast problem, we need to check whether the queue is full when enqueuing. If it is full, the enqueue operation should be notified to transfer the remaining packets to the next non-full queue for transmission.

\textbf{PCSQ enqueue:} Algorithm 1 shows the pseudocode of PCSQ enqueue function. First, the enqueue function obtains the $sid$ by parsing the SRv6 header of a packet (line 3), then it maps the $sid$ to the $qid$ based on the current transmission queue ($currQ$) state provided by the egress pipeline (line 4). If the length of the mapped queue is less than the queue size $S$, i.e., the queue is not full, the packet is enqueued  to the mapped queue and the queue length is incremented by one (line 5-7).  In most cases, the $sid$ is reasonably calculated by various scheduling algorithms in the control plane. To deal with the traffic incast problem, we allow packets with inappropriate $sid$ to shift one queue. In other words,  if the mapped queue is full and the next queue is not full, the packet can be selected to the next queue (line 8-12).  Otherwise, the packet is dropped (line 13-14). Theoretically, time-sensitive packets that cannot be discarded should enter any subsequent non-full queue. But this operation will cause persistent delay jitter and interfere with normal cyclic packet scheduling behavior. Thus, we only allow one queue shift, which keeps the jitter to one cycle.

\textbf{PCSQ dequeue and rotation:} Algorithm 2 shows the pseudocode of PCSQ dequeue and rotation function. The $t_{now}$ is the current time recorded by the RTC, and the $Flag$ is used to mark and iterate the time interval of one cycle (line 3). In one cycle time of  $C$, if the $currQ$ is not empty, the egress pipeline executes the dequeue operation and sends out all the packets in the $currQ$ (line 5-9). Since we count the length of the  enqueuing packet by adding operation, the dequeue operation can ensure that all data packets are transmitted by subtracting operation, which solves the problem of timeslot miss. After the $currQ$ is drained completely, the $t_{now}$ is updated to the RTC time again (line 10). If $t_{now}$ is still smaller than $Flag+C$, which means  the TS queue is idle, the BE queues are triggered to transmit BE packets. Otherwise, the egress pipeline executes the rotation operation and the current transmission queue is moved to the next queue (line 11-12).

\subsection{Cycle Mapping Between Nodes} \label{cmbn}

\begin{figure}[]
	\centering
		\setlength{\abovecaptionskip}{-0.4cm} 
	\includegraphics[width=0.5\textwidth]{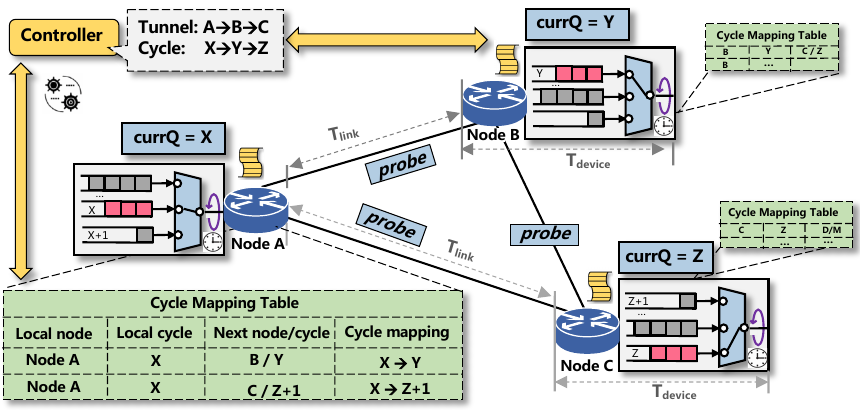}
	\caption{Cycle mapping mechanism.}
	\label{fig:cycle_mapping}
\end{figure}

Previous packet scheduling works (such as PCQ) are unaware of link delay and internal device delay. Each node is scheduled independently, which makes it difficult to deal with the aggregation problem under long-distance links. Hence, another challenge for achieving cyclic scheduling is obtaining the cycle relationship  between nodes. In a local-area network, the network-wide time synchronization can be conducted and the transmission slots can be strictly aligned (i.e., all devices in one time zone).  For a large-scale wide-area network, frequency synchronization means that devices are in different time zones. Hence, the time zone conversion is required. 

To sovle the time zone conversion problem, the cycle mapping mechanism  between nodes is proposed. For instance, there are three PCSQ-enabled nodes as shown in Figure \ref{fig:cycle_mapping}. First, when the device is initialized, the node A generates two SRv6  $probe$ packets and sends them to the adjacent node B and node C. When the  $probe$ packet is transmitted to node B at the $currQ$ of the output port,  4 bits information (i.e., 0001 to 1111 for fifteen queues) of cycle $X$  is  timestamped in the packet headers. Then, node B receives the $probe$ packet and transmits it at the $currQ$ with timestamped cycle $Y$  to node A. Thus, node A can learn the cycle mapping relationship with node B by parsing the packet header, which is $X\rightarrow Y$. In the same way, node B can learn that the relationship with node C is $Y\rightarrow Z$. Finally, each node will maintain a local cycle mapping table and update it to the control plane.  For a network with $\eta $ nodes and $\varepsilon $ edges, the total number of required $probe$ packets  is $2\left | \varepsilon \right | $, and the total number of entries for all cycle mapping tables is at most $\left | \eta \right | \times (\left | \eta \right |-1)$. Moreover,  the packet processing complexity is only $O(1)$\cite{DDR}, which is suitable for large-scale networks.

Therefore, when a tunnel of $A\rightarrow B\rightarrow C$ is conducted, the controller learns the cycle mapping relationship of $X\rightarrow Y\rightarrow Z$ in the global view, which serves as the basis for scheduling algorithms. For a naive scheduling algorithm without $sid$ computation, if the relationship is $1\rightarrow 3\rightarrow 4$, all eligible incoming packets can be naturally transmitted with cycle sequence $2\rightarrow 4\rightarrow 5$,  $3\rightarrow 5\rightarrow 6$, and so on.  


\subsection{Cycle Tags Computation} \label{ctc}

\textbf{ Programmability to approximate cyclic scheduling:} Since most cyclic scheduling algorithms (such as CQF, CSQF, Damper, and LDN) are developing, it is not trivial to implement these algorithms directly. The crucial difference is that varied constraints are used to shape the traffic and estimate the residence cycle time of packets at each hop. Thus, similar to rank computation in PIFO, we propose the concept of cycle tags computation for PCSQ to approximate a series of cyclic scheduling algorithms. Note that the value of $sid$  in Algorithm 1 is a non-negative integer determined by any customized top-level scheduling algorithm. For different time-sensitive flows, such as periodic/aperiodic flows, and constant/variable bit rate flows, various static planning or online scheduling algorithms\cite{csqf_join}\cite{csqf_huang}\cite{load_balancing_csqf} can be programmed to compute and optimize the cycle tags of packets.

\textbf{Cycle tags computation \textit{vs} rank computation:} The rank computation is restricted by local ordering that cannot guarantee the global properties such as end-to-end delay and jitter. For example, PIFO\cite{PIFO} implements the Earliest Deadline First (EDF) by using Least Slack Time First (LSTF) scheduling. The rank of a packet is computed by:
\vspace{-5pt}
$$pkt.rank = pkt.slack +pkt.arrival\_time$$
The $slack$ is the time remaining till its deadline and is decremented by the wait time at each switch’s queue. The $rank$ ensures that the packet with the closest deadline is transmitted first, but packets with specific deadlines still have the probability of being violated or even dropped. On the contrary, the cycle tags of a packet can be computed by:
$$pkt.SID = \delta = \lfloor ((pkt.deadline- (Z-X)) / h)/ T \rfloor$$
The $(Z-X)$ is the path delay measured by cycle mapping, $h$ is the number of hops, and $T$ is the cycle size. The packet's SIDs are equal to the cycle offsets $\delta$ that uniformly divides the deadline into per-hop postponed time\cite{det_serv}. Thus, the deadline is strictly satisfied by mapping flows into underlying link capacity and cycle-queue resource blocks.

Another well-investigated instance is stop-and-go\cite{SAG}, which requires packets arriving within a cycle is always transmitted at the next cycle.  PIFO defines two state variables of $cycle\_begin\_time$ and $cycle\_end\_time$ to track the beginning and end of the current cycle, where the cycle size is $T$:
\begin{equation*}
\begin{aligned}
if (now >&= cycle\_end\_time):  \\
    &cycle\_begin\_time = cycle\_end\_time  \\
   &cycle\_end\_time = cycle\_begin\_time  + T  \\
    pkt.rank& = cycle\_end\_time
\end{aligned}
\end{equation*}
In contrast, PCSQ is naturally suitable for cyclic scheduling.  We just need to set each cycle tag  to 1 (i.e., enqueue the next transmitting queue) to approximate the stop-and-go:
$$pkt.SID = \delta = 1 $$
Moreover, LDN focuses on access shaping,  thus computing the cycle tags of the first hop is enough to express LDN. Damper emphasizes reducing jitter, thus PCSQ can use a fixed cycle tags stack to simulate Damper. In fact, PCSQ's rigid scheduling is suitable for typical industrial automation traffic (such as constant bit rate traffic or flows with committed microbursts). Large burst flows are still intractable and cause a great waste of reserved resources. Apart from the queuing delay, the shaping delay of different cyclic scheduling algorithms plays an important role in the cycle tags computation. Traffic shaping at end hosts, such as Carousel\cite{Carousel} and FlowBundler\cite{FlowBundler}, can streamline the cycle tags computation by batching flows and releasing packets at scale accurately. More algorithm designs based on PCSQ will be future work.

\begin{figure*}[]
	\centering
		\subfigcapskip=-10pt
	\subfigure[ Queue size @ 10G.]{
		\begin{minipage}[t]{0.23\textwidth}
			\centering
			\includegraphics[width=\textwidth]{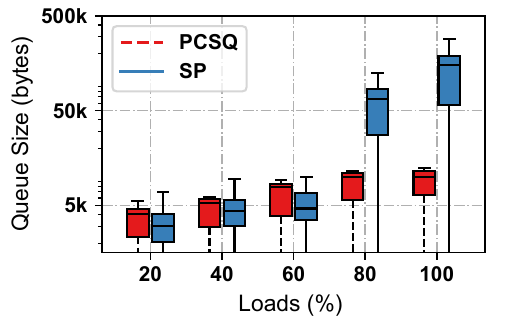}
			\label{fig1:evalutiona}
		\end{minipage}
	}
	\subfigure[ End-to-end delay @ 10G.]{
		\begin{minipage}[t]{0.23\textwidth}
			\centering
			\includegraphics[width=\textwidth]{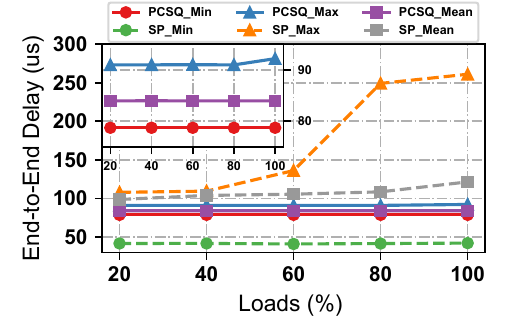}
			\label{fig1:evalutionb}
		\end{minipage}
	}
\subfigure[ Rate $vs$ delay.]{
	\begin{minipage}[t]{0.23\textwidth}
		\centering
		\includegraphics[width=\textwidth]{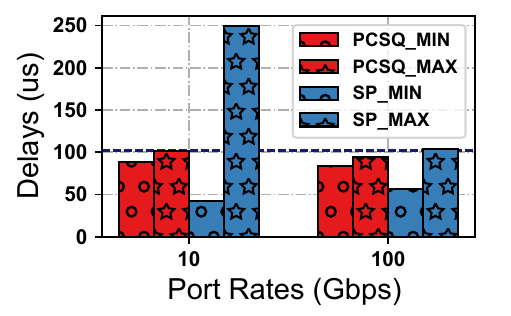}
		\label{fig1:evalutionc}
	\end{minipage}
}
\subfigure[Rate $vs$  throughput.]{
	\begin{minipage}[t]{0.235\textwidth}
		\centering
		\includegraphics[width=\textwidth]{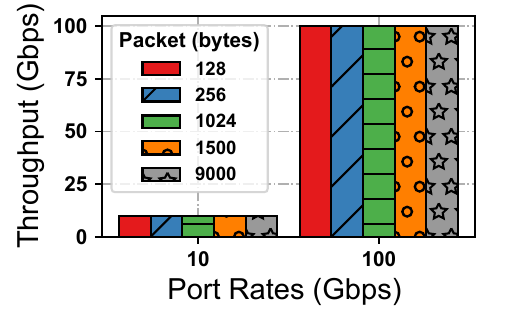}
		\label{fig1:evalutiond}
	\end{minipage}
}
	\centering
	\caption{ Evaluation results for the microbenchmark. 
	}
\end{figure*}

\begin{figure}[]
	\centering
	\subfigcapskip=-10pt
	\subfigure[ Average deviation.]{
		\begin{minipage}[t]{0.46\linewidth}
			\centering
			\includegraphics[width=\textwidth]{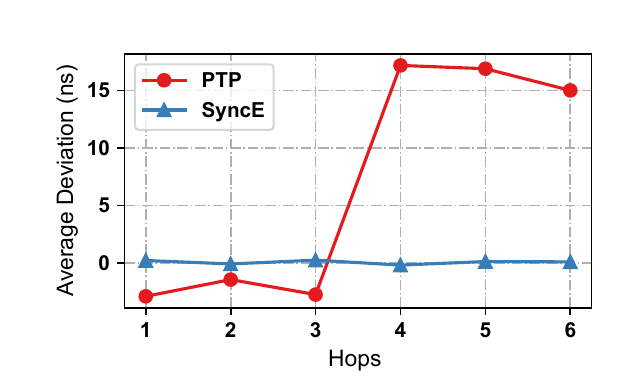}
			\label{fig2:queue_coord1}
		\end{minipage}
	}
	\subfigure[Sync waveform.]{
		\begin{minipage}[t]{0.46\linewidth}
			\centering
			\includegraphics[width=\textwidth]{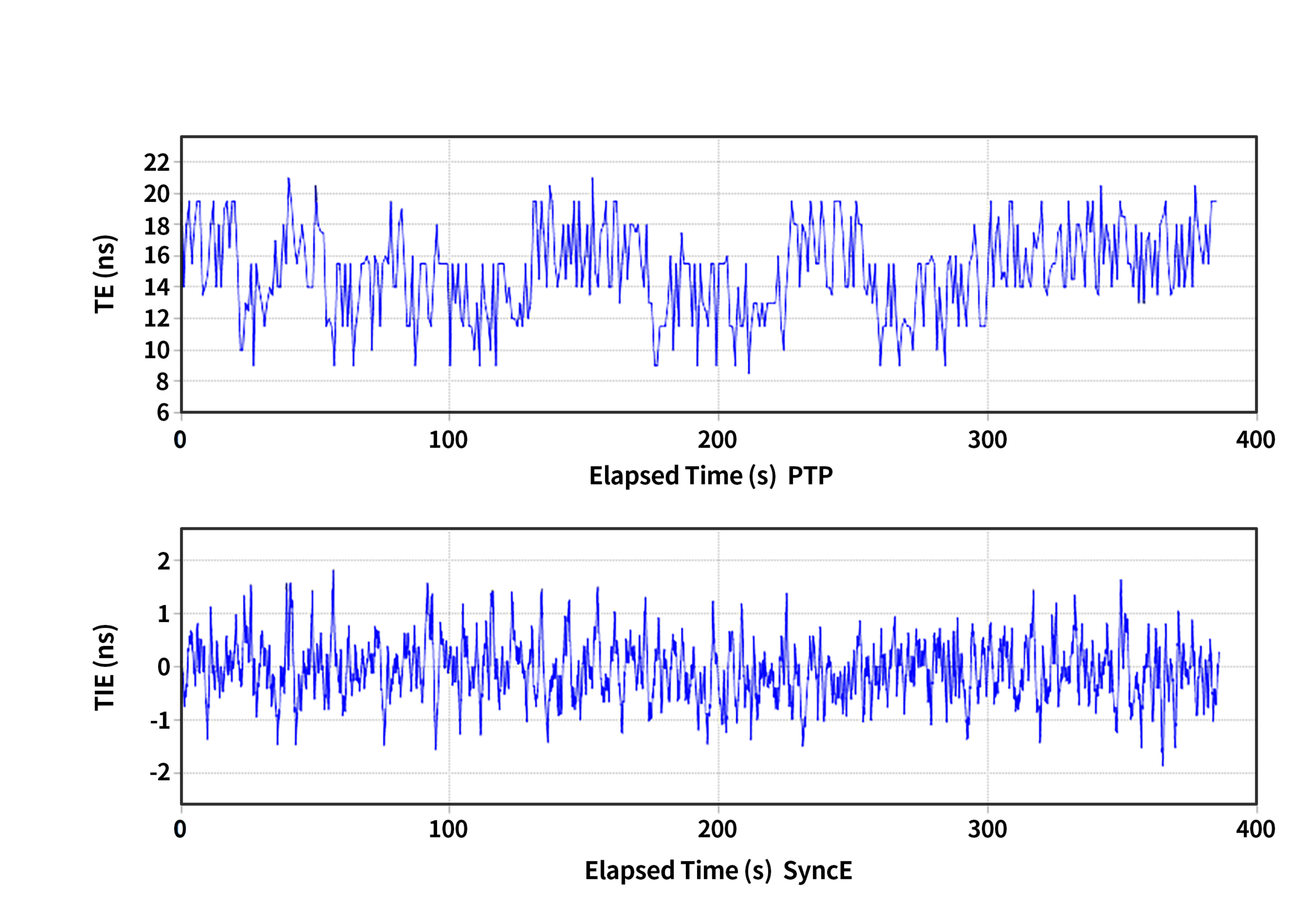}
			\label{fig2:queue_coord2}
		\end{minipage}
	}
	\centering
	\caption{ Synchronization Accuracy of SyncE and PTP.
	}
\end{figure}

\section{Evaluation}\label{evaluation}
We prototyped the PCSQ scheduler on a Xilinx KU15P FPGA\cite{ku15p} comprising 523 K CLB (Configurable Logic Blocks) LUTs (Look Up Tables), 34.6Mbits BRAM, and 10/100 Gbps interface bandwidth. Our prototype was written in System Verilog with about 1.5 K lines of code. In this section,  we first benchmark PCSQ with state-of-the-art solutions to demonstrate its performance. Besides, we evaluate PCSQ in a wide-area hardware testbed to show that PCSQ can guarantee  bounded delay and jitter transmission over long-distance links.  At last, we conduct application-level evaluations to show that PCSQ can co-transmit critical remote control flows and real-time video flows  in large-scale wide-area networks.

\subsection{Microbenchmarks}
\textbf{Evaluation setup:} We inject a large number of time-sensitive flows from multiple ingress ports to one egress port in a single PCSQ-enabled switch node. The cycle size $C$ is empirically set to 10 $\upmu$s and the queue number $N$ is set to 15. The number of queues can be larger by taking up more of the available on-chip memory. The time-sensitive flows are generated based on the token-bucket model with packet size ranges from 64 to 1500 bytes, flow rate from 256 kb/s to 10 Mb/s, and burst size from 1 to 10 packets. The maximum reserved bandwidth for time-sensitive flows is set to 12\%, and many best-effort flows are injected as background traffic. 

\textbf{Bounded latency and jitter:} First, we compare the  PCSQ with the class-based  strict priority (SP) scheduling under the port rate of 10 Gbps. As shown in Figure \ref{fig1:evalutiona}, with the loads increase, the queue length of SP scheduling grows sharply, while PCSQ maintains a strict upper bound of 12.5 KB on the queue length and enables all packets of a single cycle queue to be transmitted within 10 $\upmu$s. Then, we randomly select a time-sensitive flow to observe its end-to-end delay. In Figure \ref{fig1:evalutionb}, the minimum delay under SP scheduling is about 41.6 us, but the maximum delay reaches 261 $\upmu$s, resulting in jitter of about 219.4 $\upmu$s. The minimum and maximum delays of PCSQ are 78.7 $\upmu$s and 92.18 $\upmu$s,  which only brings 13.48 $\upmu$s of jitter. To compare the impact of different rates on the delay and jitter, we conduct the same experiment at the 100 Gbps port rate. As shown in Figure \ref{fig1:evalutionc}, the maximum delay of SP scheduling at 100G rate is reduced by 58.2\% compared with 10G rate. The increase of the port rate has little effect on the delay and jitter of PCSQ scheduling, but it will require a larger buffer to accommodate more packets under the same condition of cycle size. Additionally, PCSQ achieves more than 148 Mpps throughput, easily supporting 100 Gbps small-packet link-rate forwarding as presented in Figure \ref{fig1:evalutiond}.

\textbf{Resource overheads:} The resource usage of PCSQ is summarized in Table \ref{table1}. Compared to the WRR queue, PCSQ' URAM and BRAM consumption decreases by 12.5\% and 11.2\%. PCSQ utilizes the shared buffer to uniformly cache the deterministic flow of multiple queues into one buffer, which can not only handle the burst of each channel, but also reduce the memory resource overhead. Since the queue coordination and cycle mapping functions are implemented in LUTs and Flip Flop, PCSQ's  LUTs and Flip Flop consumption increases by 4.1\% and 3.6\% compared to FIFO queue. Moreover, PCSQ costs 0.7\% resources of DSPs to compute the timestamp and carry the timing compensation algorithm.

\setlength{\tabcolsep}{1.5mm}{
\begin{table}[h]
	\begin{tabular}{llllll}
		\toprule[1pt]
		\textbf{Resource} & \textbf{URAM}  & \textbf{BRAM}    & \textbf{LUTs}    & \textbf{Flip Flop} & \textbf{DSPs}   \\ \midrule[0.7pt]
		FIFO     & 62.5\% & 35.0\%    & 18.0\%    & 22.1\%   & 0.0\% \\
		WRR      & 62.5\% & 50.0\%    & 20.0\%    & 23.4\%   & 0.0\% \\
		PCSQ     & 50.0\%   & 38.8\% & 22.1\% & 25.7\%   & 0.7\% \\ \bottomrule[1pt]
	\end{tabular}
\caption{Summary of resource usage}
\label{table1}
\end{table}
}

\begin{figure*}[]
	\centering
		\subfigcapskip=-10pt
					\setlength{\abovecaptionskip}{-0.2cm} 
	\subfigure[ End-to-end delay.]{
		\begin{minipage}[t]{0.22\textwidth}
			\centering
			\includegraphics[width=\textwidth]{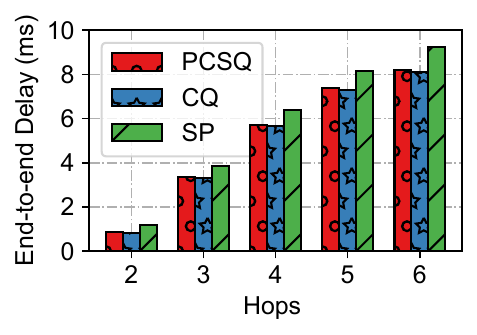}
			\label{fig3:evalutiona}
		\end{minipage}
	}
	\subfigure[ Jitter.]{
		\begin{minipage}[t]{0.28\textwidth}
			\centering
			\includegraphics[width=\textwidth]{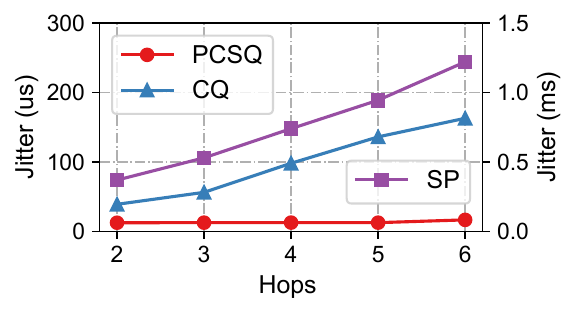}
			\label{fig3:evalutionb}
		\end{minipage}
	}
	\subfigure[ The effect of cycle size.]{
		\begin{minipage}[t]{0.22\textwidth}
			\centering
			\includegraphics[width=\textwidth]{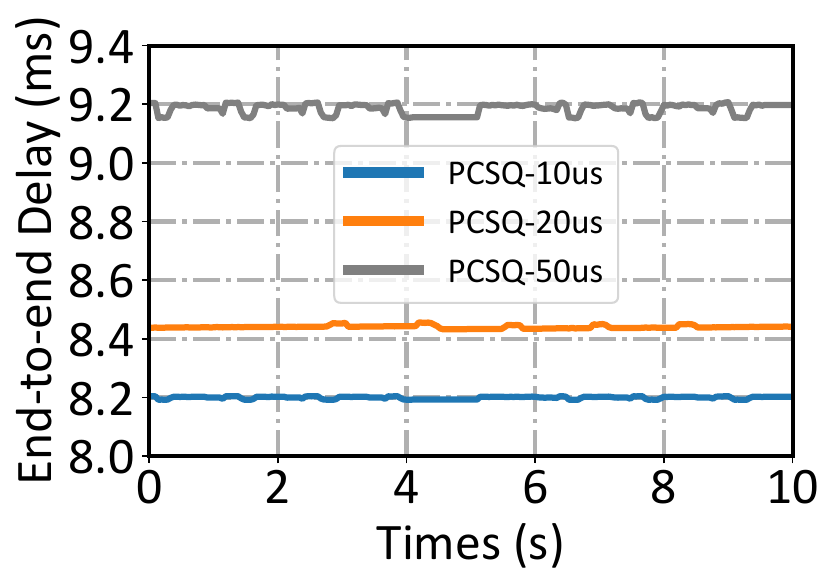}
			\label{fig3:evalutionc}
		\end{minipage}
	}
	\subfigure[Impact on best-effort flows.]{
		\begin{minipage}[t]{0.225\textwidth}
			\centering
			\includegraphics[width=\textwidth]{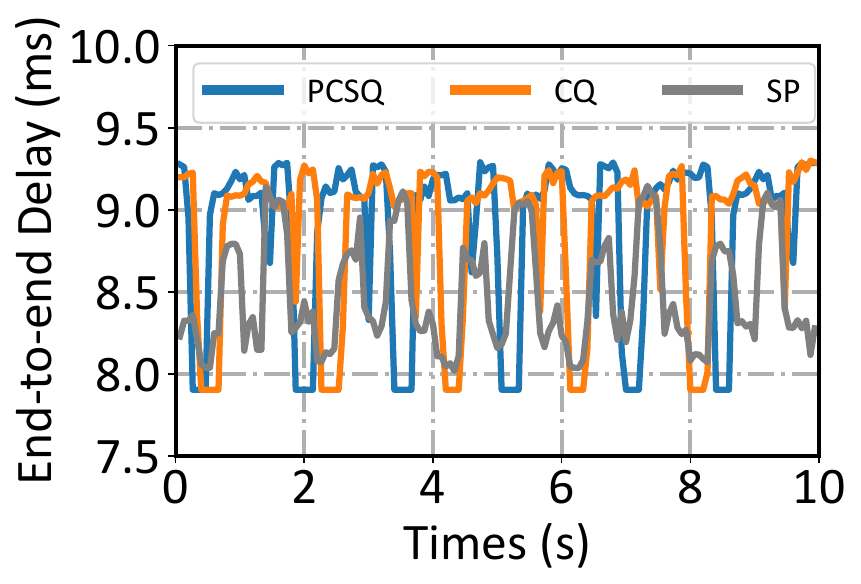}
			\label{fig3:evalutiond}
		\end{minipage}
	}
	\centering
	\caption{ Evaluation results for the long-distance testbeds. 
	}
\end{figure*}

\begin{figure}[]
	\centering
			\setlength{\abovecaptionskip}{-0.2cm} 
		\subfigcapskip=-10pt
		\begin{minipage}[t]{0.6\linewidth}
			\centering
			\includegraphics[width=\textwidth]{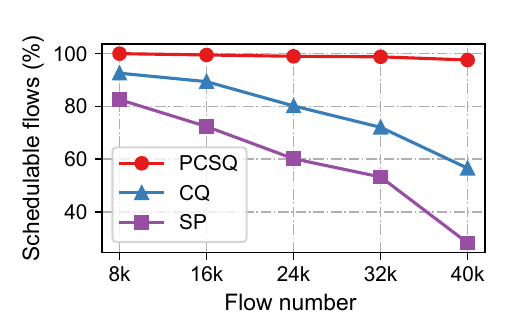}
			\label{fig5:sca_num}
		\end{minipage}
	\centering
	\caption{ The scalability of PCSQ.
	}
   \label{fig15:sca}
\end{figure}

\begin{figure*}[]
	\centering
		\subfigcapskip=-10pt
	\subfigure[ remote control flows.]{
		\begin{minipage}[t]{0.23\textwidth}
			\centering
			\includegraphics[width=\textwidth]{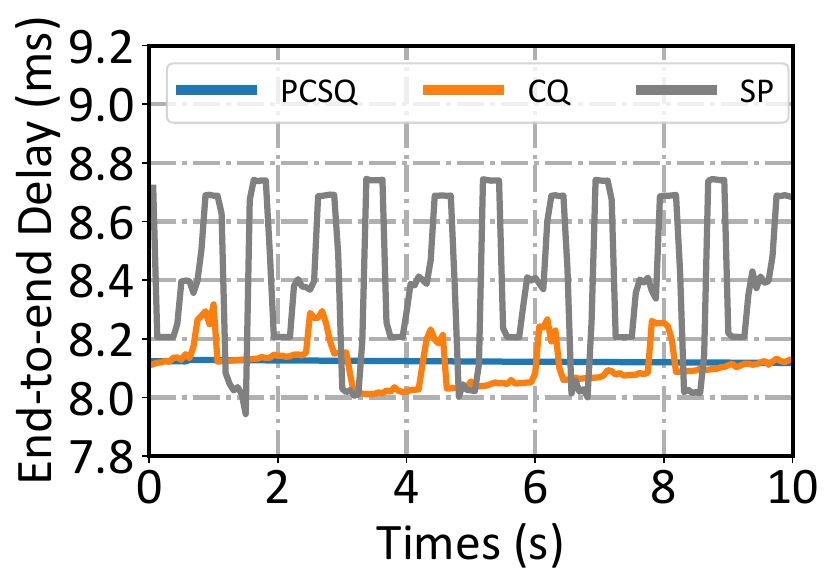}
			\label{fig4:evalutiona}
		\end{minipage}
	}
	\subfigure[ real-time video flows.]{
		\begin{minipage}[t]{0.23\textwidth}
			\centering
			\includegraphics[width=\textwidth]{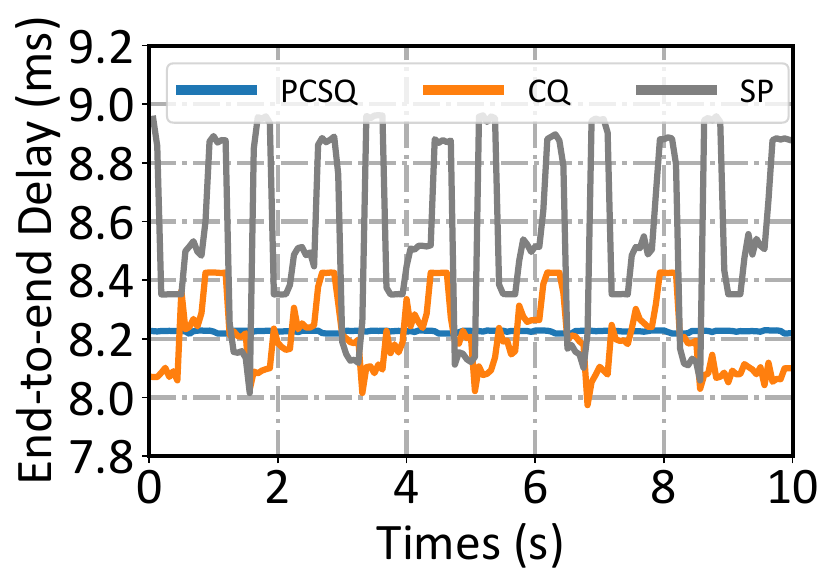}
			\label{fig4:evalutionb}
		\end{minipage}
	}
	\subfigure[ Jitter.]{
		\begin{minipage}[t]{0.25\textwidth}
			\centering
			\includegraphics[width=\textwidth]{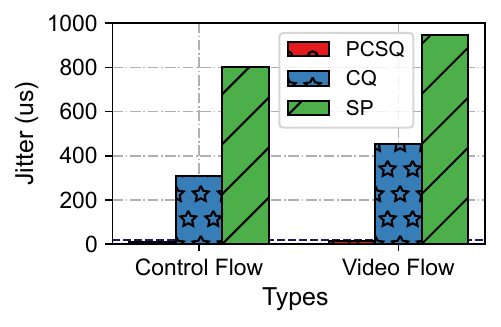}
			\label{fig4:evalutionc}
		\end{minipage}
	}
	\subfigure[Video flows as BE.]{
		\begin{minipage}[t]{0.23\textwidth}
			\centering
			\includegraphics[width=\textwidth]{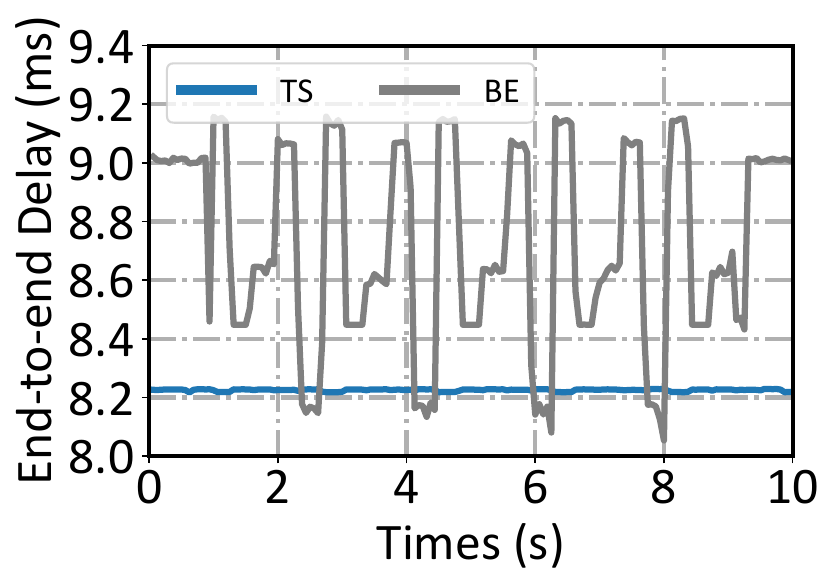}
			\label{fig4:evalutiond}
		\end{minipage}
	}
	\centering
	\caption{ Evaluation results for the industrial remote control and real-time video applications. 
	}
\end{figure*}
\subsection{Long-distance  Testbeds}
\textbf{Evaluation setup:} To evaluate the PCSQ's performance under typical long-distance aggregation scenarios, we construct a realistic testbed that contains six PCSQ-enabled nodes, which are hosted in six different cities, including Beijing, Shijiazhuang, Zhengzhou, Wuhan, Hefei, and Nanjing. The overall transmission distance from Beijing to Nanjing exceeds 1500 kilometers and the link capacity is 100 Gbps. In the experiments, the target time-sensitive flows originate from Beijing, pass through the intermediate nodes (i.e., Shijiazhuang, Zhengzhou, Wuhan, Hefei), and finally arrive at Nanjing. Numerous TS and BE traffic are injected into each node as interference flows. All flows are generated under the guidance of traffic characteristics in wide area monitoring and control systems (defined in the IEC 61850) and the industrial machine-to-machine (M2M) of the DetNet use case\cite{detnet_usecase}.

\textbf{Synchronization accuracy:} Frequency synchronization is the basis for high-precision rotation dequeuing. The PCSQ device is equipped with a  TCXO (Temperature Compensate X'tal (crystal) Oscillator) of 38.88 Mhz, and we evaluate the synchronization accuracy of the PTP protocol (1588v2 with two-way, boundary clock model) and the SyncE under the same environmental conditions. As depicted in Figure \ref{fig2:queue_coord1},  the average deviation of PTP gradually expands with the number of hops increasing. The maximum average deviation of PTP is 17.18 ns. In contrast, the average deviation of SyncE does not exceed 0.2 ns, which is stably maintained at a very low level for a long time. Figure \ref{fig2:queue_coord2} exhibits the synchronization waveform under the case of six hops. The time error of PTP ranges from 8.5 to 21 ns, while SyncE ranges from -1.85 to 1.82 ns. Based on the syncE and time compensation algorithm, PCSQ-enabled device can obtain ns-level accurate cycle time and maintain the cycle rotation consistency between adjacent nodes. 

\textbf{Performance under aggregations:} Two typical wide-area scheduling methods, strict priority (SP) and calendar queue (CQ), are set as the comparison mechanisms. CQ is a general round-robin queue that does not have enhancements such as frequency synchronization, cycle identification, and cycle mapping. Figure \ref{fig3:evalutiona} depicts the relationship between the end-to-end  worst-case delay and the number of hops. PCSQ can strictly control the upper bound of hop-by-hop network delay through the cycle tags. The delay of PCSQ is close to that of CQ, and sometimes CQ has a slightly smaller delay because packets under CQ  have some probability of entering the earlier periodic queue. Compared to SP,  PCSQ reduces the end-to-end delay by up to  10.8\%  and queuing delay by up to 70\%. More importantly, as the number of hops increases, the jitter of SP and CQ becomes larger, while PCSQ remains around 10 $\upmu$s. As shown in Figure  \ref{fig3:evalutionb}, the maximum jitter under six hops for SP, CQ and PCSQ is 1220.4 $\upmu$s, 163 $\upmu$s, and 16.7 us, respectively. Compared to SP,  PCSQ greatly reduces the jitter by up to 98.6\%. Then, we set the cycle size to 20 $\upmu$s and 50 $\upmu$s, and test the delay and jitter of the target flows under the same conditions. As shown in Figure  \ref{fig3:evalutionc}, the maximum jitter of PCSQ under cycle size of 10 $\upmu$s, 20 $\upmu$s, and 50 $\upmu$s is 12.51 $\upmu$s, 23.7 $\upmu$s, 54.45 $\upmu$s, which proves that the scheduling behavior of PCSQ conforms to the CSQF mechanism (i.e., limited to $2T$ regardless of hops). Although the cycle size of 50 $\upmu$s increases the end-to-end delay, there is no dependency between the cycle size and the delay since the delay can be reduced by adjusting the cycle tags. Furthermore, we assess the impact of PCSQ on best-effort (BE) flows. Figure  \ref{fig3:evalutiond} indicates that PCSQ will slightly increase the end-to-end delay of BE traffic compared to SP due to reserving dedicated time slots for time-sensitive traffic. To avoid starvation of BE flows, a certain percentage of gaps (e.g., 15\%) can be preset for BE flows in each cycle queue scheduled by PCSQ.

\textbf{Scalability:} To evaluate the scalability of PCSQ, we divide the time-sensitive flows into five groups of 8k, 16k, 24k, 32k, and 40k, and observe the number of schedulable flows. The deadline and jitter requirements are randomly selected from 8-10 ms and 20-500 $\upmu$s.  Figure  \ref{fig15:sca}  show that as the number of flows grows, 97.5\% of the traffic can still be successfully scheduled by PCSQ for 40k flows. Practically,  PCSQ can scale to tens of thousands of flows in large-scale deterministic networks. Since PCSQ does not need to maintain per-flow states at intermediate and egress nodes, its scheduling capability is proportional to the reserved bandwidth. As the network topology becomes larger, the complexity of cycle tag computation in the control plane may increase, but it will not affect the scheduling performance of PCSQ in the data plane. Global label optimization to maximize the schedulability will be future work.
\vspace{-10pt}
\subsection{Application-Level Performance}
In this section, we focus on specific emerging applications, such as industrial teleoperation, remote driving, remote surgery, and VR interaction. These applications often require co-transmission of remote control flows and real-time video flows. Next, we replace the target flows with the control flows and video flows, and conduct the following experiments.

\textbf{Performance of remote control flows:} The control flows are periodic mice flows, such as sending a 250-byte packet every 0.5 ms, occupying 4 Mbps of bandwidth but requiring a deadline of no more than 10 ms and jitter of 100 microseconds. Fifty control streams with different periods and packet sizes are generated and transmitted with video traffic and BE traffic under 80\% heavy-load scenarios.  As shown in Figure  \ref{fig4:evalutiona} and Figure  \ref{fig4:evalutionc},  the maximum end-to-end delay and jitter under PCSQ is 8.12 ms and 10.83 $\upmu$s, which meet the deadline  and jitter requirements for remote control.  Although the minimum delay of SP and CQ is smaller than that of PCSQ, their jitter is as high as 802.6 $\upmu$s and 309.8 $\upmu$s. 

\textbf{Performance of real-time video flows:} The single real-time video stream is quite bursty, with a mean bit rate of 5.4 Mb/s and a peak bit rate of about 120 Mb/s. A large video frame (e.g., 80 KB) is split into multiple packets not exceeding the MTU size, and large jitter between adjacent packets may cause video playback delays, quality loss, and stutters. To simulate wide-area real-time video streaming, one thousand video flows are aggregated with the mean bit rate of 5.4 Gb/s. As shown in Figure  \ref{fig4:evalutionb} and Figure  \ref{fig4:evalutionc},  the maximum end-to-end delay and jitter under PCSQ is 8.22 ms and 12.18 $\upmu$s. To keep the bursty video streams from interfering with the transmission of the control flows, we assign the two applications to different cycle queues to enhance isolation. Once the video stream does not have enough bandwidth or the cycle tag calculation is unsuccessful, it can be degraded to a best-effort stream for delivery. As shown in Figure  \ref{fig4:evalutiond}, degrading video streams will bring large latency fluctuations, while successfully tagged video streams can still maintain strictly bounded latency and jitter.
\section{Related Work}
The related packet scheduling primitives, such as PIPO and PCQ, have been elaborated in Section \ref{psd}.  There are plenty of works on  time-sensitive packet scheduling in local-area industrial networks, but few on that in wide-area networks. Next, we review some closely related ones here.

\textbf{Time-sensitive networks}: Two main local-area deterministic transmission models based on Ethernet are TTE (Time-Triggered Ethernet) and TSN. Both TTE and TSN require precise time synchronization in switches and end devices.  Most works on TSN focus on the optimization of gate control lists for single flows\cite{online_TSN}\cite{explore_limits} or multiple sub-flows\cite{tcflow}. Based on the cycle alignment\cite{ctsdn} at the edge nodes, PCSQ can connect multiple TSN domains as private WANs and realize end-to-end seamless scheduling for factory infrastructure. 

\textbf{Predictable data centers}: Cloud data centers are evolving from low latency to predictable latency\cite{aquila}\cite{fdcn}. Thanks to regular spine-leaf topology and closed control,   network calculus (such as  Silo\cite{silo}, Chameleon\cite{Chameleon}), global arbitration (e.g., Fastpass\cite{fastpass}), credit-based (e.g., ExpressPass\cite{csdb}) and window-based (e.g., $\mu $FAB\cite{ufab}) flow control, have been proposed to achieve bounded delay, low flow completion times, and fast convergence. However, all these schemes focus on the queuing delay optimization but do not consider the link delay. Combined with application-specific scheduling algorithms, PCSQ is applicable for data center interconnections.

\textbf{Deterministic networks}: There is a rough consensus that any QoS solutions that keep flow states inside the network (such as IntServ) will not scale well with the increasing demands\cite{ppv}. Thus, the wide-area deterministic schemes must be core‑agnostic or core-stateless. Dampers\cite{damper}  are presented to reduce jitter by delaying packets for the amount written in packet headers. LDN\cite{towards} scatters incoming bursts at ingress nodes and makes flows fit into the assigned cycles. PPV(Per Packet Value)\cite{ppv} provides guarantees for per-hop latency by encoding the utility function of flows to
packet value markings. These algorithms require packets to carry cycle or delay information, which all can be implemented in PCSQ.
\section{Conclusions}
We proposed a novel packet scheduler, called Programmable Cycle-Specified Queue, that can express the feature of cyclic scheduling to enable long-distance deterministic transmission. The key functional modules have been designed, including frequency synchronization, cycle  identification, queue coordination, and cycle mapping. We prototyped the PCSQ scheduler on a Xilinx KU15P FPGA and evaluated it on microbenchmarks and realistic testbeds.  The results demonstrate that PCSQ can schedule tens of thousands of time-sensitive flows and strictly guarantee ms-level delay and $\upmu$s-level jitter. 

This work does not raise any ethical issues.





\bibliographystyle{unsrt}
\bibliography{reference.bib}

\appendix

\section{Typical Scenarios and Tasks}\label{appa}
Table \ref{table2} summarizes the examples of delay, jitter, data rate, and payload size requirements for time-sensitive tasks in typical industrial scenarios, such as  discrete automation, process automation, and electricity distribution, according to 3GPP TS22261\cite{ts22261} and IETF RFC 8578\cite{detnet_usecase}. A small payload means it is less than or equal to 256 bytes, and a big payload generally does not exceed the MTU size. Note that all the values in this table are example values, which are varied in specific deployment configurations.

\begin{table}[h]
	\centering
	\scriptsize
	\begin{tabular}{p{2.3cm}p{1.2cm}p{1cm}p{1.2cm}p{1.3cm}}
		\toprule[1pt]
		Scenarios / tasks & Latency & Jitter & Data rate & Payload size \\ \midrule[0.7pt]
		
		Discrete automation& 1-10 ms & 1-100 $\upmu$s & 1-10 Mbps &  Small to big\\
		
		Process automation-remote control & 50 ms & 20 ms & 1-100 Mbps & Small to big \\
		
		Process automation-monitoring& 50 ms & 20 ms &  1 Mbps & Small \\
		
		Electricity distribution-medium voltage & 40-100 ms& 1 ms &  10 Mbps & Small to big \\
		
		Electricity distribution-high voltage & 5-10 ms & 100 $\upmu$s&  10 Mbps & Small \\
		
		Electricity distribution-extra-high voltage & 5 ms & 10 $\upmu$s & / & Small \\
		
		Intelligent transport systems- backhaul& 10 ms & 20 ms & 10 Mbps & Small to big \\
		
		Tactile interaction & 5 ms & TBC & 10 Mbps & Small to big \\
		\bottomrule[1pt]
	\end{tabular}
	\caption{Requirements for industrial tasks}
	\label{table2}
\end{table}

\section{Theoretical Analysis for Cyclic Scheduling}\label{appb}

Essentially, cyclic scheduling has a solid theoretical foundation. From the perspective of traffic characteristics, it has been proved  in\cite{flow_theory} that:
\begin{theorem} If the input flow to any network node is smooth ($s^{'}_{f}$) or uniform ($u^{'}_{f}$), then both the internal buffer and delay of that node are bounded. 
\end{theorem}

The smoothness property of a flow $s^{'}_{f}$ means that for an adjacent subsequence of fixed-size time intervals (e.g., 0-5 ms, 5-10 ms, 10-15 ms), the sum of packets within each time interval does not exceed a constant value. And uniformity is a stronger property than smoothness that the time interval is required to be narrowed to any time point. In the same vein, cyclic scheduling shapes the bursts by smoothing the traffic into different cycle-related queues with a timer to count the time expiration. And for any arbitrary time $t$ and time interval $T$, 
there is $\delta (t, T)\leq T\times C $, where the $\delta$ is the total number of incoming bits and $C$ is the link capacity. Thus, we derive the following corollary:

\begin{corollary}
	The flow shaped by cyclic scheduing is smooth flow  $s^{'}_{f}$, and the queuing delay is bounded.
\end{corollary}

\begin{figure}[]
	\centering
	\setlength{\abovecaptionskip}{-0.4cm} 
	\includegraphics[width=3.4in]{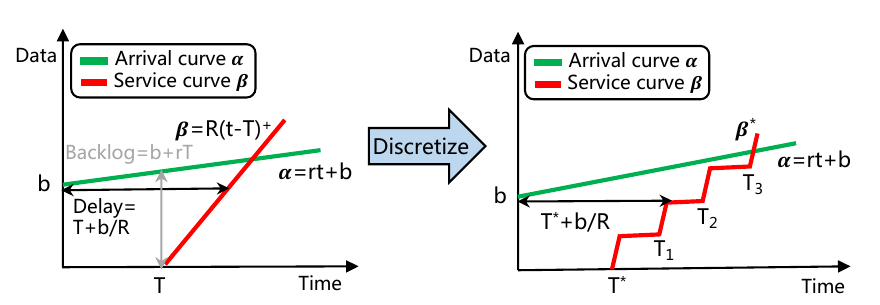}
	\caption{The cyclic scheduling can discretize the queue waiting time $T$ to $T^{*} $ in the service curve.}
	\label{fig:cal_discrete}
\end{figure}

From the perspective of the network node, cyclic scheduling can discretize the queue waiting time in the service curve. According to the network calculus theory, for a token bucket flow, the arrival curve is:
$$\alpha(t)=rt+b \text{,} \eqno{(1)}$$
where its sustainable rate is limited to $r$ B/s and bursts are up to $b$ bytes. The service curve of the network system is: $$\beta(t)=R(t-T)^{+}\text{,} \eqno{(2)}$$ 
which means data might have to wait up to $T$ seconds before being served at a rate of at least $R$ B/s. Then, the delay bound $D_{max}$ corresponds to the horizontal deviations between the arrival and service curves\cite{det_serv}, which is equal to:
$$D_{max}=T+b/R\text{.} \eqno{(3)}$$
Since the $b$ and $R$ are promised in advance, the uncertain queue waiting time (i.e., service start time) $T$ determines the size of the delay bound. As depicted in Figure \ref{fig:cal_discrete}, the cyclic scheduling weighs the queuing delay by discretizing the $T$ into $T^{*}$, which can be presented as $\left\{T_{1},T_{2}, ..., T_{n} \right\}$\cite{service_cur1}\cite{service_cur2}. Thus, packets can determine their own queue waiting time by selecting to enter different but specific cycle-related queues.

\section{CSQF Instances}\label{appc}

\begin{figure}[]
	\centering
	\setlength{\abovecaptionskip}{-0.4cm} 
	\includegraphics[width=3.4in]{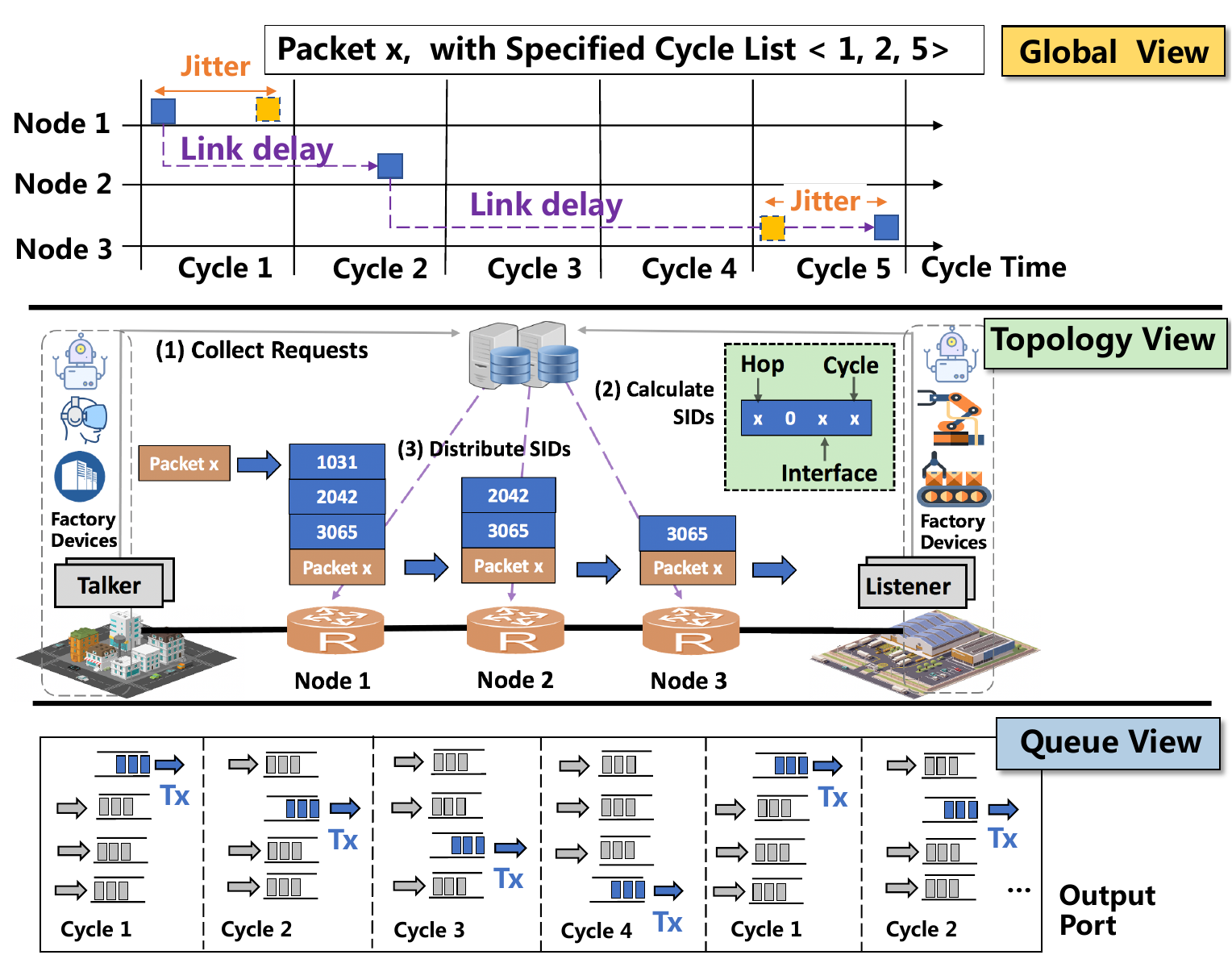}
	\caption{This is the working mechanism of CSQF in factory interconnection scenarios. For simplicity but without loss of generality, we draw frequency synchronization with the same initial phase (i.e., time synchronization) in the global view.}
	\label{fig:bg_csqf}
\end{figure}

As shown in Figure \ref{fig:bg_csqf}, CSQF attaches a list of segment routing identifiers (SIDs)\footnote{The segment routing identifiers (SIDs) are equivalent to the cycle tags.}  to a packet. The SID specifies the output interface and transmission cycle that a packet should be transmitted at each node (hop). For example, 3065 identifies cycle 5 of interface 6 at hop 3.  Moreover, CSQF adopts the frequency synchronization. Then, it enables multiple queues that one for transmitting and the remaining for receiving. Each queue corresponds to a cycle, and the transmitting queue is selected cyclically. Multiple receiving queues are used to absorb a certain amount of traffic bursts.

The workflow of connection setup is illustrated as follows: (1) A centralized controller collects the requests of QoS. (2) The controller generates the SIDs by calculating the feasible path and cycle parameters that satisfy resource and delay constraints. (3) The controller distributes the SIDs to the talker and the devices along the path. Thus, the CSQF-enabled devices can forward the packets at a precise reserved duration by consuming the first SID available in the label stack of packet headers\cite{csqf_join}.  Assuming that each output port contains two
cyclic queues, a simple but general calculation method for the maximum delay $D_{max}$
and the minimum delay $D_{min}$ is:
$$D_{max}= \sum_{i=1}^{h} (LD_{i}+PD_{i})+(h+1)T \text{,} \eqno{(2)}$$
$$D_{min}= \sum_{i=1}^{h} (LD_{i}+PD_{i})+(h-1)T \text{,} \eqno{(3)}$$
$$J_{e2e}= D_{max}-D_{min}=2T\text{,} \eqno{(4)}$$
where $LD$ is the link delay,  $PD$ is the processing delay,  $h$ is the number of hops, and $T$ is the cycle size. More importantly, since the packet can only fluctuate at the sending cycle of the first hop and the receiving cycle of the last hop, the end-to-end jitter $J_{e2e}$ is strictly limited to $2T$ regardless of network hops.


An ingress shaper can reduce the complexity of CSQF that aggregates small flows into a batch of streams and shapes traffic into desired features\cite{CENI}\cite{towards}. The ingress shaper can also convert the arrival time to the cycle time\cite{in_edge_control}\cite{towards}. Furthermore, segment routing does not need to maintain per-flow states at intermediate and egress nodes. This feature saves a significant amount of memory and helps to scale CSQF to schedule a large number of flows.

\end{document}